\documentclass[10pt]{article}

\usepackage{amsmath,amssymb,mathrsfs}
\usepackage{amsfonts}
\usepackage{bm}       
\usepackage{graphicx}
\usepackage{hyperref}
\usepackage{array}
\newcommand{\dvtx}{\mathrel{\colon}}

\usepackage{color,graphicx}
\usepackage{young}
\usepackage[vcentermath]{youngtab}
\usepackage{amsmath,amssymb,graphicx}
\usepackage{hyperref}
\definecolor{darkred}{rgb}{0.65,0.15,0}
\hypersetup{pdfborder={0 0 0},colorlinks=true,urlcolor=darkred,citecolor=blue,linkcolor=darkred,linktocpage=true}

\usepackage{cite}
\usepackage{amsmath}
\usepackage{amsfonts}
\usepackage{amssymb}
\usepackage{graphicx}%
\usepackage{amsthm}
\usepackage{mathrsfs}
\usepackage[T1]{fontenc}
\usepackage{enumerate}
\usepackage{color}

\numberwithin{equation}{section}

\title{\bf Exact solutions of the FLRW cosmological model via invariants of the Hamilton-Jacobi method}

\author{%
     E. Ahmadi-Azar%
    \thanks{\texttt{e.ahmadi.azar@azaruniv.ac.ir}},
    K. Atazadeh%
    \thanks{\texttt{atazadeh@azaruniv.ac.ir}},
    A. Eghbali%
    \thanks{\texttt{eghbali978@gmail.com}}}

\begin{document}

\maketitle

\begin{center}
    \small\emph{Department of Physics, Faculty of Basic Sciences}, \\
    \emph{Azarbaijan Shahid Madani University, 53714-161, Tabriz, Iran}
\end{center}

\vspace{1cm}
\begin{abstract}
In this study, we proceed to solve the field equations of the spatially flat Friedman-Lemaitre-Robertson-Walker (FLRW) cosmological model in the presence of the cosmological constant \(\Lambda\) by making use of the Invariants of Hamilton-Jacobi method (IHJM). This method enables us to extract systematically two independent first integrals such as \(l_{\rm HJ,1}(a,\dot{a})=c_{1}\) and \(l_{\rm HJ,2}(t,a,\dot{a})=c_{2}\) associated to the transformations group keeping the form of the Hamilton's canonical equations (HCEs) of the cosmological model invariant. Extracting these invariants means not only finding the general solution of the field equations of the model, but also obtaining the Lagrangian and Hamiltonian functions for the model whose dynamics acts like the dynamics of a single particle in a one-dimensional mini-super space \(\mathbb{Q}=(a)\). In addition, to obtain the general solution of the model, the IHJM have also solved the inverse problem of calculus of variation (IPCV) without resorting to Helmholtz conditions and whether the necessary conditions for the existence of the Lagrangian function are hold or not. The main part of the IHJM is to find the generating function of the canonical transformation (CT) and then extract two independent invariants for the desired model by using the Hamilton-Jacobi equation (HJE). This study shows that there is a close relationship between the group of the CTs of the Hamiltonian function of the particle and the one-parameter Lie group of transformations keeping invariant the Einstein-Friedmann dynamical equation (EFDE) \(\ddot{a}=F(a,\dot{a})\), so that both of them lead to the same result. In this way, having both the IHJM and the invariants of the symmetry groups method (ISGM), a comprehensive integration theory by unifying them can be achieved for studying the desired models.
\end{abstract}

\vspace{1em}

\newpage

\section{Introduction}\label{section1}

To begin the discussion, this question first arises.
What are the Noether symmetries of the FLRW cosmological model which we want to study in this article? If this cosmological model has Noether symmetry, how can the dynamical equations of the desired cosmological model be solved using these symmetries? How can the theory of ``invariants'' which was examined in the previous study \cite{1} be understood more deeply? To answer these questions, it is necessary to first see how to find a Lagrangian function for the EFDE \(\ddot{q} = F(q,\dot{q})\). The method of constructing Lagrangian and Hamiltonian structures from a constant of motion was first introduced by Sergio A. Hojman \cite{2}. Hojman's work was actually an answer to the IPCV, which is one of the famous problems in classical mechanics \cite{3}. Therefore, it is necessary to give a short history of the IPCV in this section. For this purpose, let us consider a dynamical system (DS) \(S_n^2\) with \(n\) degrees of freedom. The configuration of this system is described by an ordered \(n\)-tuple of the generalized coordinates \(q^s := (q^1, \cdots, q^n)\). Suppose the equations of motion describing the dynamics of the system \(S_n^2\) are given by the following system of second-order ordinary differential equations (ODEs):
\begin{equation}\label{Eq.(1.1)}
\ddot{q}^{k}=F^{k}(t,q^{s},\dot{q}^{s}),
\end{equation}
where \(k=1,\cdots,n\), and ``${}^{\cdot}$'' denotes the total derivative operator with respect to the time \(t\) (is the independent variable of the DS  \(S_{n}^{2}\)), which is defined as follows:
\begin{equation}\label{Eq.(1.2)}
{}^{\cdot}=\frac{{\rm d}}{{\rm d}t}:=\frac{\partial}{\partial t}+\dot{q}^{k}\frac{ \partial}{\partial q^{k}}+F^{k}\frac{\partial}{\partial\dot{q}^{k}},
\end{equation}
where \(F^{k,}\)s, \(k=1,2,...,n\), are the force functions of the DS. The IPCV can be expressed as the problem of the existence and uniqueness of a non-singular ``Hessian matrix'' \(\omega_{ik}(t,q^{s},\dot{q}^{s})\) (a square matrix of order \(n\)) and a Lagrangian function \(L(t,q^{s},\dot{q}^{s})\), which satisfy in the following relations:
\begin{equation}\label{Eq.(1.3)}
\omega_{ik}\left(\ddot{q}^{k}-F^{k}(t,q^{s},\dot{q}^{s})\right)=\frac{{\rm d}}{ {\rm d}t}\frac{\partial L}{\partial\dot{q}^{i}}-\frac{\partial L}{\partial q^ {i}}.
\end{equation}
In other words, the IPCV says whether there exists a Lagrangian function \(L(t,q^{s},\dot{q}^{s})\) and a Hessian matrix \(\omega_{ik}\), such that its corresponding equations of motion, i.e. Euler-Lagrange equations (ELEs):
\begin{equation}\label{Eq.(1.4)}
\frac{{\rm d}}{{\rm d}t}\frac{\partial L}{\partial\dot{q}^{i}}-\frac{\partial L }{\partial q^{i}}=0,
\end{equation}
are the same as the left-hand side of Eq. \eqref{Eq.(1.3)}? And if the answer is yes, is this function (Lagrangian function \( L(t,q^s,\dot{q}^s) \)) unique or multiple? The first important contributions to solve the IPCV were made by Hermann von Helmholtz \cite{4} in 1887. In 1870, Helmholtz was working on ``electrodynamics theory''. By the investigation of determining on admissible Lagrangian for Maxwell electrodynamics theory, he was brought into the research on the law of least action. He tried to generalize the variational functionals from mechanics to electrodynamics and then to some other fields of physics. In the famous treatise ``On the physical significance of the law of least action'', published in 1886, he proposed his famous conditions (which nowadays known as Helmholtz conditions) for the existence of the Lagrangian \( L \) for the IPCV \cite{5}.

\(\)
\\
\textbf{Theorem 1.1} (Helmholtz Conditions 1887). \textit{Suppose that} \(S_{n}^{2}\) \emph{be a} \emph{DS} \emph{with} \(n\) \emph{degrees of freedom whose configuration at any moment of time} \(t\) \emph{is described by the generalized coordinates} \(q^{s}(t)=\big{(}q^{1}(t),\cdots,q^{n}(t)\big{)}\). \emph{Suppose that a set of} \(n\) \emph{second-order} \emph{ODEs}:
\begin{equation}\label{Eq.(1.5)}
E_{i}(t,q^{s},\dot{q}^{s},\ddot{q}^{s}):=\omega_{ik}\left(\dot{q}^{k}-F^{k}(t,q^{s},\dot{q}^{s})\right)=0,
\end{equation}
\textit{where} \(i=1,\cdots,n\), \emph{are the governing equations of the} \emph{DS} \(S_{n}^{2}\). \emph{For this} \emph{DS}, \emph{there exist a non-singular Hessian matrix} \(\omega_{ik}\),\emph{and a Lagrangian function} \(L(t,q^{s},\dot{q}^{s})\), \emph{if and only if the Helmholtz conditions hold. According to Dauglas} \cite{6} \textit{and Sarlet} \cite{7}, \textit{these conditions are given by}:
\begin{equation}\label{Eq.(1.6)}
\omega_{ij}=\omega_{ji},\quad\frac{\partial\omega_{ij}}{\partial\dot{q}^{k}}=\frac{\partial\omega_{ik}}{\partial\dot{q}^{j}},
\end{equation}
\begin{equation}\label{Eq.(1.7)}
\mathrm{D}\big{[}\omega_{ij}\big{]}=\omega_{ik}\Gamma_{j}^{k}+\omega_{jk}\Gamma_{i}^{k},
\end{equation}
\begin{equation}\label{Eq.(1.8)}
\omega_{ik}\Phi_{j}^{k}=\omega_{jk}\Phi_{i}^{k},
\end{equation}
\textit{where}
\begin{equation}\label{Eq.(1.9)}
\Gamma_{j}^{i}:=-\frac{1}{2}\frac{\partial F^{i}}{\partial\dot{q}^{j}},\quad\Phi_{j}^{i}:=-\frac{\partial F^{i}}{\partial q^{j}}-\Gamma_{j}^{k}\Gamma_{k}^{i}-\mathrm{D}[\Gamma_{j}^{i}],
\end{equation}
\textit{and \(D\) is the total derivative with respect to the independent variable \(t\):}
\begin{equation*}
\mathrm{D}:=\frac{\mathrm{d}}{\mathrm{d}t}=\frac{\partial}{\partial t}+\dot{q}^{k}\,\frac{\partial}{\partial q^{k}}+F^{k}\,\frac{\partial}{\partial\dot{q}^{k}}.
\end{equation*}
When a solution \(\omega_{ij}\) of the system of the Helmholtz conditions, Eqs. \eqref{Eq.(1.6)}-\eqref{Eq.(1.8)} with functions \eqref{Eq.(1.9)} exists, then the
Lagrangian \(L(t,q^{s},\dot{q}^{s})\) associated to the ODE \eqref{Eq.(1.5)} is obtained from the following theorem.

\(\)
\\
\textbf{Theorem 1.2} (Engels Formula). \textit{Suppose that for the} \emph{DS} \(S_{n}^{2}\), \emph{the equations of motion are given as follows}:
\begin{equation}\label{Eq.(1.10)}
E_{i}(t,q^{s},\dot{q}^{s},\ddot{q}^{s})=0,
\end{equation}
\textit{where \(i=1,\cdots,n\). If the Helmholtz conditions} \eqref{Eq.(1.6)}-\eqref{Eq.(1.8)} \textit{with functions} \eqref{Eq.(1.9)} \textit{are satisfied by the equations of motion} \eqref{Eq.(1.10)}\textit{, the matrix \(\omega_{ij}(t,q^{s},\dot{q}^{s})\) for the} \emph{DS} \(S_{n}^{2}\) \emph{exists and the Lagrangian function} \(L(t,q^{s},\dot{q}^{s})\) \emph{can be constructed by the functions} \(E_{i}\) \emph{and} \(\omega_{ij}\), \(i,j=1,2,\cdots,n\), \emph{by using Engels formula} \cite{8,9}:
\begin{equation}\label{Eq.(1.11)}
L(t,q^s,\dot{q}^s)=\int_0^1 q^i E_i(t,\lambda q^s,\lambda \dot{q}^s,\lambda \ddot{q}^s) {\rm d}\lambda
+ \frac{{\rm d}}{{\rm d}t} \int_0^1 \int_0^1 \lambda q^i \dot{q}^j \omega_{ij}(t,\lambda q^s,\lambda \lambda' \dot{q}^s) {\rm d}\lambda\; {\rm d}\lambda'\;.
\end{equation}
\textit{Conversely, if the Lagrangian function \(L\) satisfying in} \emph{ELE}\emph{s} \eqref{Eq.(1.4)}\textit{, is given for a} \emph{DS}, \emph{then the matrix} \(\omega_{ik}\) \emph{is obtained from the following equation}:
\begin{equation}\label{Eq.(1.12)}
\omega_{ij}\bigl{(}t,q^{s},\dot{q}^{s}\bigr{)}=\frac{\partial^{2}L(t,q^{s},\dot{q}^{s})}{\partial\dot{q}^{i}\partial\dot{q}^{j}},
\end{equation}
\textit{where \(i,j=1,\cdots,n\), and this matrix satisfies the Helmholtz conditions} \eqref{Eq.(1.6)}-\eqref{Eq.(1.8)} \textit{with functions} \eqref{Eq.(1.9)}\textit{, and Engles formula} \eqref{Eq.(1.11)}. \textit{Therefore, solving the} \emph{IPCV} \emph{means to find the matrix} \(\omega_{ik}\) \textit{which satisfies in the Helmholtz conditions.}
\({}\)

In 1894, G. Darboux was completely solved IPCV in one dimension. He showed that in one-dimensional case, the Lagrangian function for a second-order differential equation usually exists and is non-unique \cite{10}. Since the publication of Helmholtz's paper (1887) until the beginning of the 20th century, the IPCV has been examined from differential aspects. For example, Mayer \cite{11} proved the part of Helmholtz sufficient condition. In 1901, Knigsberger \cite{12} proposed a comprehensive and detailed review of the IPCV in general case. In 1903, Hamel \cite{13} solved a special case of IPCV in three dimensions. After that, no important work was done in this field until Douglas \cite{6} in 1941 in a detailed article titled ``Solution of the inverse problem of calculus of variations'', solved the inverse problem for the two-dimensional case. His method was mainly based on the use of the theory of partial differential equations (PDEs) about the differential system related to the desired DS. After this article, no important research was done on the IPCV until the late seventies of the 20th century. In the late 70s and 80s, the IPCV was noticed again. During this time, physicists followed two different approaches in investigating the subject. In the first approach, the DS was studied from the perspective of the geometry of the manifolds. In this method, the problem of symmetries and constants of motion was also studied from a geometrical point of view, and the transformation relations of equivalent Lagrangians were also proposed in the same way. In the second approach, the working basis was the Helmholtz condition. These conditions were algebraic-differential equations, and solving this system of equations means finding the matrix $\omega_{ik}(t,q^s,\dot{q}^s)$ and the Lagrangian function $L(t,q^s,\dot{q}^s)$ and thus solving the IPCV for the desired system \cite{7}. In the late 1980s and early 1990s, many studies were expanded and led to results in other branches of physics and even mathematics. The clear example of these researches was Havas's studies \cite{14,15}. Havas made progress toward the solution of IPCV for the first-order differential equation, which was completely solved by Hojman and Urrutia \cite{16}. They provided a method for constructing infinitely many Lagrangian functions for such system of differential equations \cite{17,18}. As was mentioned earlier, in one-dimensional case, Darboux proved that Lagrangians usually exists, but two constants of motion are needed to construct these Lagrangians. In 1996, Hojman proved with new methods that only one constant of motion and one symmetry transformation are needed to constructed the Lagrangian function. Later, Hojman completed his studies. In 2014, he showed that to construct a Lagrangian for a system of two first-order ODEs or one second-order ODE, it is enough to know only one time-independent constant of motion. He presented an elegant way to construct the Lagrangian function for these DSs where only one time-independent constant of motion is known. One can find a full review of the IPCV and also full account of the latest developments in Refs. \cite{19,20,21}.

The paper is organized as follows. After Introduction section,  Section \ref{section 2} summarizes the main points of the FLRW cosmological.
In this section, from the Einstein's field equations (EFEs) in the presence of the cosmological constant for the spatially
flat FLRW cosmological model, we first derive Friedman equations and then show that the dynamics of the
FLRW cosmological model behaves like single-particle dynamics in which a one-dimensional force is applied to it.
In Section \ref{section 3}, we will use the Hojman's method to construct the Lagrangian and Hamiltonian functions for the DS of the FLRW cosmological model.
Then, by using the Rand-Trautman identity, we will obtain the Noether point symmetries of the dynamical equation (DE). In addition, in this section, by using the Noether theorem, we will extract two invariants (first integrals) for the DE of the FLRW cosmological model.
In Section \ref{section 4}, we apply the CT in Hamilton-Jacobi theory for the FLRW cosmological model with the Hamiltonian function \eqref{Eq.(3.42)} to extract the invariants associated to the EFDE \eqref{Eq.(2.9)}.
In this way, one gets the corresponding invariants. Finally, at the end of this section by using the Hamilton-Jacobi invariants we obtain the general solution of the EFDE \eqref{Eq.(2.9)} from solving the HJE \eqref{Eq4.23}.
We conclude with a final discussion of the results with remarks and perspectives.
Some of the definitions and theorems applied in the text of paper such as Poisson and Lagrange brackets matrices, Hojman's formula, etc are given in Appendix A.
Also, in Appendix B we introduce the CT in Hamilton-Jacobi theory to discuss the method of obtaining the general solutions of system of HCEs and HJE.

\section{FLRW cosmological model}\label{section 2}

Before proceeding to study the Hojman's method for deriving the Lagrangian functions for the Friedmann DEs in cosmology, it is necessary to derive the Friedmann equations in the spatially flat (\( k = 0 \)) FLRW cosmological model. For this cosmological model, the general relativity (GR) is chosen as the background theory. We must first select a suitable metric tensor for space-time, which represents the symmetry of the universe with a good approximation, that is, homogeneity and isotropy. This metric is given by \cite{22}
\begin{equation}\label{Eq.(2.1)}
{\rm d}s^2 = -c^2{\rm d}t^2 + a^2(t)\Big(\frac{{\rm d}r^2}{1-kr^2}+r^2{\rm d}\theta^2+r^2\sin^2\theta\;{\rm d}\varphi^2\Big),
\end{equation}
where in the space-time coordinates \( x^\mu := (t, r, \theta, \varphi) \), \( t \) is the cosmic time (\( t \geq 0 \)) and the variables \( 0 \leq r < \infty \), \( -\pi \leq \theta \leq \pi \), and \( 0 \leq \varphi \leq 2\pi \) are the spherical polar coordinates, and \( a(t) \) is a differentiable function of the cosmic time which is called the cosmological scale factor, \( k \) is the curvature constant of three-dimensional space which can accept three values \(-1, 1, 0\) are related to the open, closed, and flat world, respectively.

Now, let us have a short discussion about the underlying theory of GR. One of the ways to deduce classical fields, including the equations of the GR, is to use ``the principle of least action''. This principle says that the nature of all the possible geometries for the space-time manifold M, accepts a geometry where the gravitational action in the presence of the matter, which is called the Einstein-Hilbert action \cite{32}
\begin{equation}\label{Eq.(2.2)}
S_{\rm EH} := \frac{1}{2\kappa c} \int_{\Omega} (R + 2\Lambda + 2\kappa c \mathcal{L}_{\rm M}) \sqrt{-g} \, \rm d^4x,
\end{equation}
is an extreme value, that is, its variation with respect to the dynamical variable \( g_{\mu\nu} \) is equal to zero: \( \delta S_{\rm EH} = 0 \), where \( \delta \) denotes a small variation. In action
\eqref{Eq.(2.2)}, \( R \) is the Ricci scalar, \( \Lambda \) is the cosmological constant, \( g = \text{det} g_{\mu\nu}, \sqrt{-g} \, \rm d^4x \) is the volume invariant of the space-time manifold \( \rm M, \kappa (=8\pi G/c^4) \) is the Einstein's gravitational constant, and \( \Omega \) is a region of the space-time manifold where the gravitational field is present. By varying the Einstein-Hilbert action \( S_{\rm EH} \), \eqref{Eq.(2.2)},  with respect to the metric tensor \( g_{\mu\nu} \), yields the EFEs \cite{32}
\begin{equation}\label{Eq.(2.3)}
R_{\mu\nu} - \frac{1}{2} g_{\mu\nu} R + \Lambda g_{\mu\nu} = \kappa T^{\rm M}_{\mu\nu},
\end{equation}
where \( T^{\rm M}_{\mu \nu} \) is the energy-momentum tensor of the matter defined in terms of the Lagrangian density \( \mathcal{L}_{\rm M} \), such that \cite{32}
\begin{equation}\label{Eq.(2.4)}
T^{\rm M}_{\mu \nu} = \left( \rho + \frac{p}{c^2} \right) U_\mu U_\nu + p g_{\mu \nu},
\end{equation}
where \( U_\mu \)'s are the covariant components of the four-velocity vector \( U = U^\mu \partial_\mu = (c, 0, 0, 0) \) of an observer co-moving with the fluid. Furthermore, the pressure \( p \) and density of fluid \( \rho \) are related together via the equation of state \( p = w \rho c^2 \), where the coefficient \( w \) is a real dimensionless number which is called the parameter of the equation of state. In this study, our metric signature is \( (-+++) \) and we use the natural units where the velocity of light in vacuum is unity: \( c = 1 \).

Now, we are in position to derive the Friedmann equations in the spatially flat (\( k = 0 \)) FLRW cosmological model with the line element \eqref{Eq.(2.1)}.
In the presence of the cosmological constant \( \Lambda \), Eq. \eqref{Eq.(2.3)} for the line element \eqref{Eq.(2.1)} leads to \cite{42}
\begin{equation}\label{Eq.(2.5)}
3\frac{\dot{a}^2}{a^2} = 8\pi G \rho + \Lambda,
\end{equation}
\begin{equation}\label{Eq.(2.6)}
-2\frac{\ddot{a}}{a} - \frac{\dot{a}^2}{a^2} = 8\pi G p - \Lambda.
\end{equation}
To determine three unknowns \( a(t) \), \( \rho(t) \), and \( p(t) \), we have only two independent equations \eqref{Eq.(2.5)} and \eqref{Eq.(2.6)}. Hence, in order to be able to find these unknowns, we also need an equation, and it is nothing but equation of state \( p = w\rho \). This algebraic equation together with two differential equations \eqref{Eq.(2.5)} and \eqref{Eq.(2.6)} form a system of three differential-algebraic equations for three unknowns \( a(t) \), \( \rho(t) \) and \( p(t) \). The combination of the three equations of this system leads to the following non-linear second-order ODE:
\begin{equation}\label{Eq.(2.7)}
\ddot{a} = \alpha \frac{\dot{a}^2}{a} + \beta a,
\end{equation}
where the new parameters \( \alpha \) and \( \beta \) are defined as follows:
\begin{equation}\label{Eq.(2.8)}
\alpha := -\frac{1 + 3w}{2}, \quad \beta := \frac{(1 + w)\Lambda}{2}.
\end{equation}
In order to solve Eq. \eqref{Eq.(2.7)} , we change the dependent variable \( a(t) \) to the new variable \( q(t) \). In this way, Eq. \eqref{Eq.(2.7)}  can be written as
\begin{equation}\label{Eq.(2.9)}
\ddot{q} = F(q, \dot{q}),
\end{equation}
where
\begin{equation}\label{Eq.(2.10)}
F(q, \dot{q}) := \alpha \frac{\dot{q}^2}{q} + \beta q.
\end{equation}
Eq. \eqref{Eq.(2.9)}  in one-dimensional configuration space \( \mathbb{Q} = (q) \) (the mini-super space), is the Newton's second law, where the function \eqref{Eq.(2.10)} , is the component of the force acting on a particle of unit mass in this space. Note that the FLRW cosmological model in the framework of the GR theory can be imagined as the dynamics of a particle in one dimension, which is subjected to the force
\begin{equation}\label{Eq.(2.11)}
\textbf{F} = F(q, \dot{q}) \frac{\partial}{\partial q},
\end{equation}
and the particle under the influence of this force has acceleration \(\ddot{q}\). The first step to solve Eq. \eqref{Eq.(2.9)}  in the mini-super space \( \mathbb{Q} = (q) \) via the IHJM, is to find the Lagrangian and Hamiltonian functions corresponding to the particle in the mini-super space. We will do this in the next section.

\section{Construction of Lagrangian and Hamiltonian Structures}\label{section 3}

In order to construct a Lagrangian function $L(t,q,\dot{q})$ for the DS of the FLRW cosmological model, which has an EFDE following Eq. \eqref{Eq.(2.9)},
we use the Hojman's method \cite{2,16,44}.
First, we define the 2-tuples $x^{k}:=(x^{1},x^{2}):=(q,\dot{q})$ as the generalized coordinates of the particle and $(\Phi^{1},\Phi^{2}):=(\dot{q},F(q,\dot{q}))$ as the force acting on the particle in this new two-dimensional space. Then, one can write the system of differential equations \eqref{Eq.(2.9)}  in the following form
\begin{equation}\label{3.1}
\frac{\mathrm{d}x^{k}}{\mathrm{d}t}=\Phi^{k}(x^{i}).
\end{equation}
where $k=1,2$.
Now, let us consider a DS of consisting a particle with $n$, ($n=1,2,\cdots$), degrees of freedom. Suppose that $\mathbb{Q}$ be the configuration space of this particle and let $q^{s}:=$ $(q^{1},\cdots,q^{n})$ be the local generalized coordinates at the arbitrary point $P\in\mathbb{Q}$. The DEs of the particle on the configuration space $\mathbb{Q}=q^{s}=(q^{1},\cdots,q^{n})$ read
\begin{equation}\label{3.2}
\dot{q}^{\alpha}=F^{\alpha}(t,q^{s},\dot{q}^{s}).
\end{equation}
where the force functions $F^{\alpha}$'s, $\alpha=1,\cdots,n$, are differentiable functions of the generalized coordinates $q^{s}:=(q^{1},\cdots,q^{n})$, the generalized velocities $\dot{q}^{s}:=({\dot{q}}^{1},\cdots,{\dot{q}}^{n})$, and time $t$ (the independent variable). By defining a single set of $2n$ variable $x^{i}:=$ $(q^{s},{\dot{q}}^{s})$ as the generalized coordinates and a single set of $2n$ variables $\Phi^{i}:=({\dot{q}}^{s},F^{s})$ as the force functions acting on the particle in the new tangent bundle $\mathrm{T}\mathbb{Q}$, the DEs on the configuration space $\mathbb{Q}=$ $q^{s}$, can be transformed to a set of $2n$ first-order ODEs on the $\mathrm{T}\mathbb{Q}$ of the following form \eqref{3.1}. The first $n$ of these equations read $\mathrm{d}q^{\alpha}/\mathrm{d}t=\dot{q}^{\alpha},\ \alpha=1,\cdots,n$, and the last are the DEs \eqref{3.2}. Accordingly,
the equations governing the particle motion in the two-dimensional tangent bundle \(\mathrm{T}\mathbb{Q}\), are the following first-order ODEs:
\begin{equation}\label{3.3}
\dot{x}^{1}=\Phi^{1}\big{(}x^{k}\big{)}=\dot{q}=x^{2},
\end{equation}
\begin{equation}\label{3.4}
\dot{x}^{2}=\Phi^{2}\big{(}x^{k}\big{)}=\alpha\frac{\dot{q}^{2}}{q}+\beta q= \alpha\frac{(x^{2})^{2}}{x^{1}}+\beta x^{1}.
\end{equation}
By using the equations of motion \eqref{3.3} and \eqref{3.4}, one can find the force acting on the particle in this manifold
\begin{equation}\label{3.5}
    \mathbf{\Phi} = \Phi^i \frac{\partial}{\partial x^i}
    = x^2 \frac{\partial}{\partial x^1}
    + \left( \alpha \frac{(x^2)^2}{x^1} + \beta x^1 \right) \frac{\partial}{\partial x^2}.
\end{equation}
In order to apply the Hojman's method to the DS \(S^{2}_{1}\), we need a time-independent invariant. This invariant can be easily extracted by doing a little calculation using the symmetry. The DE of the DS \(S^{2}_{1}\) is considered to be the EFDE \(\Gamma(q,\dot{q},\ddot{q}):=\ddot{q}-F(q,\dot{q})=0\).
This equation is not an explicit function of time.
So, it is possible to reach one of the Lie group of transformations without using the Lie invariance condition.
This group of transformations is the following time translation group \(\mathbf{T}_{\varepsilon}\colon\mathbb{R}^{+}\times\mathbb{R}\to\mathbb{R}^{+}\times\mathbb{R}\), where
\begin{equation}\label{3.6}
\begin{array}{c}
(t,q)\mapsto\big{(}\bar{t},\bar{q}\big{)}=\mathbf{T}_{\varepsilon}(t,q)=(t+\varepsilon,q),\end{array}
\end{equation}
and \(\mathbb{R}^{+}=[0,+\infty)\). The DE \(\Gamma=0\) remains invariant under the ``one-parameter Lie group of transformation'' \eqref{3.6}. The symmetry vector \(\mathbf{X}=\partial_{t}\), where \(\partial_{t}:=\partial/\partial t\), is the infinitesimal generator of this group of transformation. The existence of this transformation group indicates that in the DS \(S^{2}_{1}\), the energy of the particle is conserved. Therefore, one of the first integrals of the DS is the energy of the particle. We denote this invariant by the symbol \(I(q,\dot{q})\). To find \(I(q,\dot{q})\), we solve the EFDE \(\Gamma=0\) by the method of separation of variables. By using the chain rule, equation \(\Gamma=0\) can be written as follows:
\begin{equation}\label{Eq.(3.7)}
\Gamma=\dot{q}\frac{\mathrm{d}\dot{q}}{\mathrm{d}q}-\alpha\frac{\dot{q}^{2}}{q}-\beta q=0.
\end{equation}
The presence of the term \(-\beta q\) into Eq. \eqref{Eq.(3.7)} makes it impossible to separate the variables \(q\) and \(\dot{q}\) and solve the differential equation by the method of the separation of variables. Therefore, we leave this term. With the other two terms, we define a new differential equation in the form
\begin{equation}\label{Eq.(3.8)}
\Gamma^{*}=\dot{q}\frac{\mathrm{d}\dot{q}}{\mathrm{d}q}-\alpha\frac{\dot{q}^{2}}{q}=0.
\end{equation}
Suppose that the first integral of this equation is a function such as \({C}_{1}(q,\dot{q})\). To find this, one must solve equation \(\Gamma^{*}=0\) by the method of separation of variables.
Then, we have \(\mathrm{d}\dot{q}/\dot{q}=\alpha\mathrm{d}{q}/q\). By integrating, we obtain \(\ln \dot{q}=\alpha\ln q+C_{2}\). This equation gives the integration constant \(C_{2}\) as follows: \(C_{2}=\ln(\dot{q}/q^{\alpha})\). Here, we choose \(C_{1}=e^{2C_{2}}/2\), which is an invariant for the differential equation \(\Gamma^{*}=0\), as the first integral.
Therefore, equation \(C_{1}=e^{2C_{2}}/2\) can be written as \(C_{1}(q,\dot{q})=\dot{q}^{2}q^{-2\alpha}/2\). It should be noted that since \(C_{2}\) is a first integral, any differentiable function of this first integral also will be a first integral.
Thus, the sum of two first terms of the differential equation \(\Gamma=0\) gives us the kinetic energy and then the third term \(-\beta q\) definitely gives the potential energy of the particle.
Now, to find the invariant \(I(q,\dot{q})\), associated to the ODE \eqref{Eq.(3.7)}, let us consider the following ansatz for the function \(I(q,\dot{q})\) as \(I(q,\dot{q})=C_{1}(q,\dot{q})+C_{3}(q)\), where the unknown function \(C_{3}(q)\) is the first integral associated to the third term of \(\Gamma=0\). In order to find the function \(C_{3}(q)\), we take the total derivative from both sides of Eq. \(I(q,\dot{q})=C_{1}(q,\dot{q})+C_{3}(q)\), with respect to time \(t\):
\begin{equation}\label{Eq.(3.9)}
\dot{q}q^{-2\alpha}\big(\ddot{q}-\alpha\frac{\dot{q}^{2}}{q}+\frac{1}{\dot{q}q^{-2\alpha}}\frac{\mathrm{d}C_{3}(q)}{\mathrm{d}t}\big)=0. \end{equation}
In order Eq. \eqref{Eq.(3.9)} to always holds for all values of the variable \emph{q} in its domain, it is necessary and sufficient that the expression inside the parenthesis is equal to zero, i.e.,
\begin{equation}\label{Eq.(3.10)}
\ddot{q}-\alpha\frac{\dot{q}^{2}}{q}+\frac{1}{\dot{q}q^{-2\alpha}}\frac{\mathrm{d}C_{3}(q)}{\mathrm{d}t}=0.
\end{equation}
By comparing \eqref{Eq.(3.10)} with that of \eqref{Eq.(3.7)}, we get \({\rm d}C_{3}=-\beta q^{-2\alpha+1}{\rm d}q\), ant then \(C_{3}(q)=Cq^{2-2\alpha}/2\), where \(C:=\beta/(1-\alpha)\) is a new parameter.
Hence, the time-independent invariant of \eqref{Eq.(3.7)} is
\begin{equation}\label{Eq.(3.11)}
I(q,\dot{q})={1\over 2}\dot{q}^{2}q^{-2\alpha}-{1\over 2}Cq^{-2\alpha+2},
\end{equation}
which gives the energy function of the particle in the DS \(S_{1}^{2}\).
This time-independent invariant extraction method can be used for many DEs that are not an explicit function of time.
Now that the time-independent invariant associated to  Eq. \eqref{Eq.(2.9)} has been found, we are in a position to apply the Hojman's method to the DS \(S_{1}^{2}\).
Our goal in this section is to derive the Lagrangian and Hamiltonian functions for the DS \(S_{1}^{2}\).
Having the invariant \eqref{Eq.(3.11)} as a time-independent constant of motion for the particle, we can define a Hamiltonian structure for \(S_{1}^{2}\) including a Hamiltonian function and a Poisson Brackets relation in terms of the antisymmetric matrix \(J^{ij}\) \cite{2,48,49,50,51} (see Appendix A).

Now let us return to our problem. In order to find the Poisson Brackets matrix \( J^{ij}(x^k) \) and
the Lagrange Brackets matrix \( \sigma_{ij}(x^k) \) for a particle in the DS \( S_2^1 \) with a time-independent constant of motion \( I(x^k) \),
given by Eq. \eqref{Eq.(3.11)}, we use the Hamilton's equations \eqref{Eq.(3.15)} and the components of the force \(\mathbf{\Phi}\) of Eq. \eqref{3.5}.
Then, we find that
\begin{equation}\label{Eq.(3.20)}
J^{ij}(x^k) =
\begin{pmatrix}
0 & (x^1)^{2\alpha} \\
-(x^1)^{2\alpha} & 0
\end{pmatrix},
\quad \sigma_{ij}(x^k) =
\begin{pmatrix}
0 & (x^1)^{-2\alpha} \\
-(x^1)^{-2\alpha} & 0
\end{pmatrix}.
\end{equation}
It should be noted that the matrices in \eqref{Eq.(3.20)} satisfy Eq. \eqref{Eq.(3.19)}.
Now we are in a position to write the EFDE of the FLRW cosmological model \eqref{Eq.(2.9)} in terms of Poisson and Lagrange Brackets matrices \( J^{ij} \) and \( \sigma_{ij} \) in the DS \( S_2^1 \). For this purpose, combining the Hamilton equations \eqref{Eq.(3.15)}, the equations of motion \eqref{3.2}, and then substituting the invariant \( I(x^k) \) into the Hamiltonian function \( H \), we obtain the following equations:
\begin{equation}\label{Eq.(3.21)}
\dot{x}^i = J^{ij}\frac{\partial I(x^k)}{\partial x^j}, \quad i = 1, 2.
\end{equation}
To express these equations in terms of the Lagrange Brackets matrix, one must multiply both sides of Eq. \eqref{Eq.(3.21)} by the matrix \( \sigma_{ki} \) and then employ Eq. \eqref{Eq.(3.19)}.
The result is
\begin{equation}\label{Eq.(3.22)}
\sigma_{ij} \dot{x}^j + \frac{\partial I(x^k)}{\partial x^i} = 0, \quad i = 1, 2.
\end{equation}
Thus, for the DS \( S^1_2 \), if the time-independent invariant \( I(x^k) \) and the Lagrange Brackets matrix \( \sigma_{ij} \) are known, then Eq. \eqref{Eq.(3.22)} enables us another way to express the equation of motion of the particle in the tangent bundle \( \mathrm{T}\mathbb{Q} = (q, \dot{q}) \). According to Eq. \eqref{Eq.(3.17)}, we have
\begin{equation}\label{Eq.(3.23)}
L(x^k) = l_1(x^k) \dot{x}^1 - \frac{1}{2}(x^2)^2(x^1)^{-2\alpha} + \frac{1}{2}C(x^1)^{-2\alpha+2},
\end{equation}
where \( l_1(x^k) \) as the general solution of the ODE \eqref{Eq.(3.18)} is \( l_1(x^k) = (x^1)^{-2\alpha}x^2 \).
Then, plugging this into Eq. \eqref{Eq.(3.23)} we obtain the Lagrangian function in the following form
\begin{equation}\label{Eq.(3.24)}
L(x^k) = \frac{1}{2}(x^2)^2(x^1)^{-2\alpha} + \frac{1}{2}C(x^1)^{-2\alpha+2}.
\end{equation}
Hence, by considering 2-tuple \( (x^1, x^2) = (q, \dot{q}) \) in \eqref{Eq.(3.24)}, the Lagrangian function of the particle in the tangent bundle \(\mathrm{T}\mathbb{Q}\) becomes
\begin{equation}\label{Eq.(3.25)}
L(q, \dot{q}) = \frac{1}{2} \dot{q}^2 q^{-2\alpha} + \frac{1}{2} C q^{-2\alpha+2} \in \mathcal{F}(\mathrm{T}\mathbb{Q}).
\end{equation}
One may define the kinetic and potential energy functions corresponding to the Lagrangian \eqref{Eq.(3.25)}
as \(T(q, \dot{q}) = \dot{q}^2 q^{-2\alpha}/2\) and \(V(q) = -C q^{-2\alpha+2}/2,\), respectively.
To transition from the Lagrangian formalism of the DSs to the Hamiltonian formalism, it is necessary to first define the generalized momentum.
\( \)
\\
\textbf{Definition 3.1} (Generalized Momentum, Hojman 2014). \emph{For the particle in the DS} \(S_{2}^{1}\) \emph{with one degree of freedom, the configuration space} \(\mathbb{Q} = (q)\), \emph{and the Poisson Brackets matrix} \(J^{ij}(x^{k})\), \emph{i}, \emph{j} = 1, 2, \emph{the generalized momentum canonically conjugate to the generalized coordinate} \(x^{1} = q\), \emph{is defined as follows}:
\begin{equation}\label{Eq.(3.26)}
p(x^{k}) := \int \frac{dx^{2}}{J^{12}(x^{k})}.
\end{equation}
Indeed, this definition is consistent with the formal definition \(p := \partial L / \partial \dot{x}^{1}\), that is,
\begin{equation}\label{Eq.(3.27)}
p(x^{k}) = \frac{\partial L(x^{k})}{\partial \dot{x}^{1}} = \int \frac{dx^{2}}{J^{12}(x^{k})}.
\end{equation}
The generalized momentum \eqref{Eq.(3.27)} does not represent the components of a vector on the configuration space \(\mathbb{Q}\) but rather a co-vector. The 2-tuple \((q, p)\) is to be thought of not as local coordinates in the tangent bundle but as coordinates for ``cotangent bundle''. Eq. \eqref{Eq.(3.27)} is then to be considered as the local description of a map \( p : \mathrm{T}\mathbb{Q} \rightarrow \mathrm{T}^{*}\mathbb{Q}\)
\begin{equation}\label{Eq.(3.28)}
(q, \dot{q}) \mapsto p(q, \dot{q}) = \Big( q = q, p = \int \frac{dx^{2}}{J^{12}(x^{k})} \Big),
\end{equation}
from the tangent bundle to the cotangent bundle. The map \eqref{Eq.(3.28)} is, in fact, the formal definition for the generalized momentum of
canonically conjugate to the generalized coordinate \( x^1 = q \). Having the Poisson brackets matrix \( J^{ij} \) and by calculating the
integral \eqref{Eq.(3.26)}, the generalized momentum \( p \) can be obtained as a function of the variables \( q \) and \( \dot{q} \).
In this map, the coordinate \( x^1 = q \) should be kept unchanged. It should be noted that the 2-tuple \( (q,p) \) is the local
coordinates for \( \mathrm{T}^{*}\mathbb{Q} \). The space \( \mathrm{T}^{*}\mathbb{Q} \) consists of co-vectors to configuration space
which is usually called in the classical mechanics, the ``phase space'' of the DS \( S_1^2 \).
The \( p  \) is the component of a 1-form field \(\boldsymbol{\omega} = p\,\mathrm{d}q\),
not the component of a vector field. The 1-forms are dual to the vector fields, so \( \boldsymbol{\omega} \)
lies in a space that dual to \( \mathrm{T}_q\mathbb{Q} \). This space is denoted by \( \mathrm{T}^{*}\mathbb{Q} \) and called the cotangent space at the point \( q \in \mathbb{Q} \).

Now we have to prepare ourselves to be able to construct the basic function of the Hamiltonian dynamics, i.e., the Hamiltonian function.
For this purpose, we first discuss the ``Legendre transformation'' (see, Appendix \ref{app.A}). We use this transformation to derive the HCEs. As we known, one of the simplest and most useful contact transformations is the Legendre transformation. Contact transformations differ from the point transformations in that the functions defining the transformations in addition of the coordinate dependence, also depend on the derivatives of the dependent variable. The Legendre transformation has some remarkable properties and provides the link between the Lagrangian and Hamiltonian formalisms of the DSs. The Legendre transformation allows us to construct the Hamiltonian function from the known Lagrangian function of the given DS. As mentioned earlier, the Lagrangian function is defined on the tangent bundle of the configuration space \( \mathrm{T}\mathbb{Q} \), while the Hamiltonian function is defined on the its dual \( \mathrm{T}^{*}\mathbb{Q} \).

In this section, we found the Lagrangian function for the FLRW cosmological model as \eqref{Eq.(3.25)}. Now, for the tangent bundle \( \mathrm{T}\mathbb{Q} = (x^1, x^2) \) corresponding to the configuration space \( \mathbb{Q} = (x^1) = (q) \), we define a phase space (or cotangent bundle) \( \mathrm{T}^{*}\mathbb{Q} = (x^1, p) \) by the map \( \text{\textbf{FL}: } \mathrm{T}\mathbb{Q} \to \mathrm{T}^{*}\mathbb{Q} \), where:
\begin{equation}\label{Eq.(3.40)}
(x^1, x^2) \mapsto \text{\textbf{FL}}(x^1, x^2) = (x^1 = x^1, p = l_1(x^1, x^2)),
\end{equation}
and \( l_1(x^1, x^2) = x^2(x^1)^{-2\alpha} \). Conversely, any point of the cotangent bundle \( \mathrm{T}^{*}\mathbb{Q} = (x^1, p) \) transforms into a point of the tangent bundle \( \mathrm{T}^*\mathbb{Q} = (x^1, x^2) \) by the inverse of the map \( \text{\textbf{FL}} \), that is, map \( (\text{\textbf{FL}})^{-1} : \mathrm{T}^{*}\mathbb{Q} \to \mathrm{T}\mathbb{Q} \), where
\begin{equation}\label{Eq.(3.41)}
(x^1, p) \mapsto (\text{\textbf{FL}})^{-1}(x^1, p) = (x^1 = x^1, x^2 = l_{2}(x^1, p)),
\end{equation}
where \( l_2 \) is a function of variables \( x^{1} \) and \( p \). Notice that the transformation equation \( x^2 = l_2(x^1, p) \) in the map \eqref{Eq.(3.41)}, is the solution of the transformation equation \( p = l_1(x^1, x^2) \) for the variable \( x^2 \) in terms of the variables \( x^1 \) and \( p \). The transformation equations of the map \( (\text{\textbf{FL}})^{-1} \) in \eqref{Eq.(3.41)} are: \( x^1 = x^1,\) and \(x^2 = p(x^1)^{2\alpha} \).

Now, let us consider the time-independent invariant Eq. \eqref{Eq.(3.11)}, which is the energy function of the particle in the DS \(S^{1}_{2}\). Substituting the 2-tuple \((x^{1},x^{2})=(q,\dot{q})\) in this equation and then using the transformation equation \(x^{2}=p(x^{1})^{2\alpha}\) in the resulting equation, yields us the desired invariant in the cotangent bundle (or phase space). The resulting quantity is called the Hamiltonian function of the particle on the phase space, which is denoted by \(H(q,p)\):
\begin{equation}\label{Eq.(3.42)}
H(q,p)=\frac{1}{2}p^{2}q^{2\alpha}-\frac{1}{2}Cq^{-2\alpha+2}\in\mathcal{F}( \mathrm{T}^{*}\mathbb{Q}).
\end{equation}
In this way, the FLRW cosmological model with the EFDE \eqref{Eq.(2.9)} can be thought of as one-dimensional particle dynamics.
The generalized coordinate and the generalized velocity of this particle are respectively \(q\) and \(\dot{q}\).
The one-dimensional force acting on the particle which depends on the position \(q\) and velocity \(\dot{q}\) is given by Eq. \eqref{Eq.(2.11)}.
The Lagrangian and Hamiltonian functions of this particle are given by Eqs. \eqref{Eq.(3.25)} and \eqref{Eq.(3.42)}, respectively.
The Lagrangian function \(L(q,\dot{q})\) is defined on the space \(\mathcal{F}(\mathrm{T}\mathbb{Q})\) of the tangent bundle \(\mathrm{T}\mathbb{Q}=(q,\dot{q})\), while the Hamiltonian function \(H(q,p)\) is defined on the space \(\mathcal{F}(\mathrm{T}^{*}\mathbb{Q})\) of the cotangent bundle \(\mathrm{T}^{*}\mathbb{Q}=(q,p)\). Now, to investigate that these functions fulfilled into \eqref{Eq.(3.33)}, let us assume that \(L\) is known, then the Jacobi condition  \eqref{Eq.(3.33)} holds, \(\det\omega_{11}(q,\dot{q})=q^{-2\alpha}\neq 0\).
The Jacobi condition \eqref{Eq.(3.33)} means that the necessary condition of the {Theorem A.2} holds, and therefore, the FLRW cosmological model is a hyper-regular Lagrangian system. Then, according to the {Theorem A.2}, there is a Hamiltonian function for this DS.
According to Eq. \eqref{Eq.(3.34)}, this function is obtained as follows: \(H=p\dot{q}-L(q,\dot{q})\), where \(\dot{q}(q,p)\) is the solution of the equation \(p=\partial L/\partial\dot{q}\). According to Eq. \eqref{Eq.(3.36)}, the generalized momentum canonically conjugates to the generalized coordinate \(q\) is as follows:
\begin{equation}\label{Eq.(3.43)}
p=\frac{\partial}{\partial\dot{q}}\Big(\frac{1}{2}\dot{q}^{2}q^{-2\alpha}+\frac{1}{2}Cq^{-2\alpha+2}\Big)=\dot{q}q^{-2\alpha}. \end{equation}
Solving Eq. \eqref{Eq.(3.43)} for the variable \(\dot{q}\) in terms of the canonical variables \((q,p)\), gives the relation \(\dot{q}=pq^{2\alpha}\). By plugging this into \(H=p\dot{q}-L\), we thus get the Hamiltonian function \eqref{Eq.(3.42)}.
This is exactly the function that we obtained earlier by the Hojman's method, where in this method the Poisson Brackets matrix is used.

Now one question arises. Is it possible to formulate the equations of motion of the particle in DS \(S^{2}_{n}\) with the configuration space \(\mathbb{Q}:=q^{s}=(q^{1},\cdots,q^{n})\) by means of the Hamiltonian function \(H(q^{s},p_{s})\) instead of the Lagrangian function \(L(q^{s},\dot{q}^{s})\)?
To answer this question, we take a partial derivative in both sides of Eq. \eqref{Eq.(3.35)}, with respect to the fibre \(p_{\alpha}\)'s, \(\alpha=1,\cdots,n\):
\begin{equation}\label{Eq.(3.44)}
\frac{\partial H}{\partial p_{_\alpha}}=\frac{\partial}{\partial p_{_\alpha}}\left(p_{_\beta}\dot{q}^{\beta}-L\right)=\dot{q}^{\alpha}.
\end{equation}
Once again, we take partial derivative from both sides of Eq. \eqref{Eq.(3.35)}, with respect to the coordinate \(q^{\alpha}\):
\begin{equation}\label{Eq.(3.45)}
\frac{\partial H}{\partial q^{\alpha}}=\frac{\partial}{\partial q^{\alpha}}\left(p_{_\beta}\dot{q}^{\beta}-L\right)=-\dot{p}_{\alpha},
\end{equation}
where Eq. \eqref{Eq.(3.36)} and ELEs \eqref{Eq.(1.4)} are used. Therefore, Eqs. \eqref{Eq.(3.44)} and \eqref{Eq.(3.45)} provide the basic equations of the Hamiltonian dynamics, which are called the HCEs. These equations only include the Hamiltonian function \(H(q^{s},p_{s})\) and canonical variables \((q^{s},p_{s})\). It should be noted that the system of Eqs. \eqref{Eq.(3.44)} and Eqs. \eqref{Eq.(3.45)}
form \(2n\) first-order ODEs for \(2n\) unknowns \((q^{s},p_{s}):=(q^{1},\cdots,q^{n},p_{1},\cdots,p_{n})\). In Hamiltonian formalism, these first-order HCEs are replaced by the system of \(n\) second-order ELEs \eqref{Eq.(1.4)} for \(n\) unknown \(q^{s}:=(q^{1},\cdots,q^{n})\). These two sets of ODEs, ELEs and HCEs, are completely equivalent provided that the Jacobi condition \eqref{Eq.(3.33)}
is satisfied. One of the main advantages of the Hamiltonian formalism is that the \(q^{\alpha}\)'s and \(p_{\alpha}\)'s, are considered on an equal footing. In order to do this consistently, let us consider the set of \(2n\) variables \(\xi^{k}\)'s, \(k=1,\cdots,2n\), which are defined as follows \cite{48}:
\begin{equation}\label{Eq.(3.46)}
\xi^{k}=\left\{\begin{array}{ll}q^{k},&k=1,\cdots,n,\\ p_{_{k-n}},&k=n+1,\cdots,2n,\end{array}\right.
\end{equation}
where, the first \(n\) of the \(\xi^{k}\)'s in Eq. \eqref{Eq.(3.46)}, are the generalized coordinates \(q^{s}=(q^{1},\cdots,q^{n})\), and the second \(n\) are the generalized momentums \(p_{s}=(p_{1},\cdots,p_{n})\) of the particle. Now, we are in position to prove that the HCEs \eqref{Eq.(3.44)} and \eqref{Eq.(3.45)} can be expressed in a unified form. For this purpose, we define \(2n\) variables \(Q^{k}\)'s:
\begin{equation}\label{Eq.(3.47)}
Q^{k}:=\left\{\begin{array}{ll}\dfrac{\partial H}{\partial p_{_k}}=\dfrac{\partial H}{\partial\xi^{k+n}},&k=1,\cdots,n,\\\\ -\dfrac{\partial H}{\partial q^{k}}=-\dfrac{\partial H}{\partial\xi^{k-n}},&k=n+1,\cdots,2n,\end{array}\right.
\end{equation}
as force functions acting on the particle in the cotangent bundle \({\rm T}^{*}\mathbb{Q}\). According to definition \eqref{Eq.(3.47)}, the first \(n\) of the \(Q^{k}\)'s are \(\partial H/\partial\xi^{k+n}\)'s, and the last \(n\) are \(-\partial H/\partial\xi^{k-n}\)'s. By using Eqs. \eqref{Eq.(3.46)} and \eqref{Eq.(3.47)}, the HCEs \eqref{Eq.(3.44)} and \eqref{Eq.(3.45)} in terms of the unified coordinates \(\xi^{k}\) and unified force \(Q^{k}\) become: \(\dot{\xi}^{k}=Q^{k}\)'s, \(k=1,\cdots,2n\). Now, let us define the \(2n\times 2n\) ``symplectic matrix'' S with the elements \(S^{jk}\)'s, \(j,k=1,\cdots,2n\), as follows:
\begin{equation}\label{Eq.(3.48)}
S^{jk}=\begin{pmatrix}O_{n}&I_{n}\\ -I_{n}&O_{n}\end{pmatrix},
\end{equation}
where \(I_{n}\) and \(O_{n}\) are the \(n\times n\) unit and null matrices, respectively.
Using \eqref{Eq.(3.48)}, the HCEs \(\dot{\xi}^{k}=Q^{k}\)'s, \(k=1,\cdots,2n\), can be written on the cotangent bundle \({\rm T}^{*}\mathbb{Q}\) as follows \cite{46}:
\begin{equation}\label{Eq.(3.49)}
\dot{\xi}^{k}=S^{kj}\dfrac{\partial H (\xi^i)}{\partial\xi^{j}},
\end{equation}
where \(k=1,\cdots,2n\) and \(S^{jk}\in\mathcal{F}({\rm T}^{*}\mathbb{Q})\).
Eq. \eqref{Eq.(3.49)} is called the unified form of the HCEs \eqref{Eq.(3.44)} and \eqref{Eq.(3.45)}.
The similarity of Eq. \eqref{Eq.(3.49)} in Hamiltonian formalism with Eqs. \eqref{Eq.(3.21)} in Hojman formalism is remarkable. Having Eq. \eqref{Eq.(3.21)}, to write the unified form \eqref{Eq.(3.49)}, it is enough to change \(x^{i}\to\xi^{i}\), \(\dot{q}^{s}\to p_{s}\), \(J^{ij}\to S^{ij}\), \(I\big{(}x^{i}\big{)}\to H\big{(}\xi^{i}\big{)}\), where \(i,j=1,\cdots,2n\), \(s=1,\cdots,n\), and \(I(x^{i})\in\mathcal{F}({\rm T}\mathbb{Q})\) is the energy function (time-independent invariant) of the particle in the tangent bundle \({\rm T}\mathbb{Q}\) of the configuration space \(\mathbb{Q}=q^{s}:=(q^{1},\cdots,q^{n})\).

Let us apply Eq. \eqref{Eq.(3.49)} for the DS of the FLRW cosmological model. It should be remembered that the Hamiltonian function for the particle in the FLRW cosmological model is given by Eq. \eqref{Eq.(3.42)}.
The symplectic matrix \(S^{ij}\) in this model is a \(2\times 2\) matrix. Using this and then
substituting the Hamiltonian function \eqref{Eq.(3.42)} into the HCEs \eqref{Eq.(3.49)} we obtain
\begin{equation}\label{Eq.(3.51)}
\dot{q}=\dot{\xi}^{1}=S^{12}\frac{\partial H}{\partial\dot{\xi}^{2}}=pq^{2\alpha},
\end{equation}
\begin{equation}\label{Eq.(3.52)}
\dot{p}=\dot{\xi}^{2}=S^{21}\frac{\partial H}{\partial\dot{\xi}^{1}}=- \alpha p^{2}q^{2\alpha-1}+\beta q^{-2\alpha+1}.
\end{equation}
Combining Eqs. \eqref{Eq.(3.51)} and \eqref{Eq.(3.52)} one can get
\begin{equation}\label{Eq.(3.53)}
\frac{\mathrm{d}}{\mathrm{d}t}\left(\dot{q}q^{-2\alpha}\right)=-\alpha(\dot{q}q^{-2\alpha})^{2}q^{2\alpha-1}+\beta q^{-2\alpha+1}.
\end{equation}
Taking the derivative of left-hand side of Eq. \eqref{Eq.(3.53)} with respect to time and then simplifying the resulting equation, the EFDE \eqref{Eq.(2.9)} is obtained.

Now, for the DS \(S_{1}^{2}\) with one degree of freedom and the Lagrangian function \eqref{Eq.(3.25)}, we define functional action \(S[q(t)]\) by the following finite integral:
\begin{equation}\label{Eq.(3.54)}
S[q(t)]=\int_{t_{1}}^{t_{2}}\left(\frac{1}{2}\,\dot{q}^{2}\,q^{-2\alpha}+\frac{1}{2}\,C\,q^{-2\alpha+2}\right)\mathrm{d}t\,.
\end{equation}
In fact, the functional action \eqref{Eq.(3.54)} is a map from a set of functions to the set of real numbers \({\mathbb{R}}\). The domain of the map is a set of twice differentiable functions \(\Omega=\{q(t)\colon t\in[t_{1},t_{2}]\}\) on the closed interval \([t_{1},t_{2}]\). This map is given by \cite{60} \(S\colon\Omega\to{\mathbb{R}}\), where
\begin{equation}\label{Eq.(3.55)}
q(t)\mapsto S[q(t)]=\int_{t_{1}}^{t_{2}}L\,\left(t,q(t),\dot{q}(t)\right)\mathrm{d}t.
\end{equation}
\\
\textbf{Definition 3.2} (Invariance of the Functional Action). \emph{Suppose that for the DS} \(S_{1}^{2}\) \emph{with one degree of freedom, and the Lagrangian function} \(L\left(t,q,\dot{q}\right)\), \emph{the finite integral in the map} \eqref{Eq.(3.55)} \emph{be a functional action on the closed interval} \([t_{1},t_{2}]\subset{\mathbb{R}}^{+}\). \emph{This functional action under the one-parameter Noether group of point transformation} \cite{61,62,63} \(\mathbf{\Phi}_{\varepsilon}^{\rm(N)}\colon{\mathbb{R}}^{+}\times{\mathbb{R}} \to{\mathbb{R}}^{+}\times{\mathbb{R}}\), where
\begin{equation}\label{Eq.(3.56)}
(t,q)\mapsto(\bar{t},\bar{q})=\mathbf{\Phi}_{\varepsilon}^{\rm(N)}(t,q)=\big{(}\alpha(t,q;\varepsilon),\psi(t,q;\varepsilon)\big{)},
\end{equation}
\emph{is called invariant, whenever there is a scalar field such as} \(N(t,q;\varepsilon)\) \emph{that satisfies in the following relation}:
\begin{equation}\label{Eq.(3.57)}
\dot{\alpha}L\left(\alpha,\psi(t,q;\varepsilon),(\dot{\alpha})^{-1}\dot{\psi}(t,q;\varepsilon)\right)=L\big{(}t,q(t),\dot{q}(t)\big{)}+\frac{\mathrm{d}N}{\mathrm{d}t}\,.
\end{equation}
\emph{This relation is called ``the invariance of the functional action condition'' or the variational symmetry condition} \cite{62}.

\(\)

It should be noted that for a given Lagrangian function, if the variational symmetry condition \eqref{Eq.(3.57)} holds, then the DS has ``variational symmetry''.
In such DS, to find the first integrals (or invariants) we must use Noether's theorem. This theorem for DSs consisting of particles with one degree of freedom can be presented as follows:

\(\)
\\
\textbf{Theorem 3.1} (Noether's Theorem). \emph{Suppose that for the DS} \(S^{2}_{1}\) \emph{with one degree of freedom and with Lagrangian function} \(L\left(t,q,\dot{q}\right)\), \emph{the functional action} \(S[q(t)]\) \emph{under the one-parameter Noether group of point transformation} \eqref{Eq.(3.56)} \emph{with the infinitesimal transformation equations}

\[t\rightarrow\bar{t}=\alpha(t,q;\varepsilon)=t+\varepsilon\tau(t,q)+\mathcal{O}(\varepsilon^{2}),\]
\begin{equation}\label{Eq.(3.58)}
q\rightarrow\bar{q}=\psi(t,q;\varepsilon)=q+\varepsilon\xi(t,q)+\mathcal{O}(\varepsilon^{2}),
\end{equation}
\emph{is invariant, that is, the variational symmetry condition}  \eqref{Eq.(3.57)} \emph{holds, then there is a function such as}
\begin{equation}\label{Eq.(3.59)}
I(t,q,\dot{q})=-\frac{\partial L}{\partial\dot{q}}\left(\xi-\dot{q}\tau\right)-L\tau+G,
\end{equation}
\emph{called the Noether charge (or Noether invariant), which satisfies in the Noether identity:}
\begin{equation}\label{Eq.(3.60)}
\left(\xi-\dot{q}\tau\right)\frac{\delta S}{\delta q}=\frac{{\rm d}I}{{\rm d}t}\cdot
\end{equation}
\emph{The Noether charge along any solution of the equation of motion of the DS \(S^{2}_{1}, \delta S/\delta q=0\), is conserved, i.e.,}
\begin{equation}\label{Eq.(3.61)}
\frac{{\rm d}I}{{\rm d}t}\Big|_{\frac{\partial S}{\partial q}}=0.
\end{equation}
\emph{In the Noether charge}  \eqref{Eq.(3.59)}, \(G:=\partial N/\partial\varepsilon\big|_{\varepsilon=0}\) \emph{is called the Bessel-Hagen (BH) term or gauge term or even sometimes the boundary term}.
\\
\textbf{Proof}. See Refs. \cite{61,62,68}.

\( \)

The Noether identity can be rewritten as follows:
\begin{equation}\label{Eq.(3.62)}
\left(\xi-\dot{q}\tau\right){\rm E}(L)=\frac{{\rm d}}{{\rm d}t}\left(G+H\tau-p\xi\right),
\end{equation}
where
\begin{equation}\label{Eq.(3.63)}
{\rm E}:=\frac{\partial}{\partial q}-\frac{{\rm d}}{{\rm d}t}\frac{\partial}{\partial\dot{q}},
\end{equation}
is the Lagrangian operator (or variational derivative), and \( E(L) = \delta S / \delta q = 0 \) is the ELE. Sometimes, Eq. \eqref{Eq.(3.62)} is called the ``Noether-Bessel-Hagen Identity'' (NBHI) \cite{69}.
It should be noted that the functions \( \tau(t,q) = \partial \alpha / \partial \varepsilon |_{\varepsilon=0} \)
and \( \xi(t,q) = \partial \psi / \partial \varepsilon |_{\varepsilon=0} \) in Eqs. \eqref{Eq.(3.58)} are the infinitesimals of the Noether point symmetry \eqref{Eq.(3.56)}.
For a given Lagrangian function, dealing with this invariance criterion, Eq. \eqref{Eq.(3.57)} is not a suitable for the study of the Noether point symmetries of problem, because solving Eq. \eqref{Eq.(3.57)} to find the infinitesimals of the Noether point symmetry can be difficult. Hence, around 1970, H. Rund and A. Trautman to solve this problem, presented a useful theorem for the invariance criterion of a functional action \cite{63}. If one differentiates in both sides of Eq. \eqref{Eq.(3.57)} with respect to the parameter \( \varepsilon \), and then set \( \varepsilon = 0 \), the resulting equation will be the Rand-Trautman identity, which is repressed as follows \cite{64}:

\( \)
\\
\textbf{Theorem 3.2} (Rand-Trautman Identity). \emph{Suppose that for a DS} \( S^2_1 \) \emph{with the Lagrangian function} \( L(t,q,\dot{q}) \), \emph{the finite integral}
\begin{equation}\label{Eq.(3.64)}
S[q] = \int_{t_1}^{t_2} L(t,q,\dot{q}) \mathrm{d}t,
\end{equation}
\emph{be the functional action on the closed interval} \([t_1,t_2] \subset \mathbb{R}^+ = (0,+\infty)\). \emph{If this functional action is invariant under the one-parameter Noether group of point transformation}  \eqref{Eq.(3.56)} \emph{with the infinitesimal transformation equations}  \eqref{Eq.(3.58)}, \emph{then the infinitesimals} \( \tau(t,q) \) and \( \xi(t,q) \) \emph{are satisfied in the Rand-Trautman identity}:
\begin{equation}\label{Eq.(3.65)}
\text{RT}(L,\mathbf{X},G) := \dot{G} - \mathbf{X}^{[1]} (L) - \dot{\tau} L = 0,
\end{equation}
\emph{where the vector field} \( \mathbf{X} = \tau \partial_t + \xi \partial_q \) \emph{is the generator of} \( \mathbf{\Phi}^{(\mathrm{N})}_\varepsilon \), \emph{and} \( \mathbf{X}^{[1]} \) \emph{is the first-order prolongation of} \( \mathbf{X} \) \emph{which is defined as follows}:
\begin{equation}\label{Eq.(3.66)}
\mathbf{X}^{[1]} := \tau \frac{\partial}{\partial t} + \xi \frac{\partial}{\partial q} + (\dot{\xi} - \dot{\tau}\dot{q}) \frac{\partial}{\partial \dot{q}}.
\end{equation}

In general, for a DS \( S^2_n \) with \( n \) degrees of freedom and the Lagrangian function \( L(t,q^s,\dot{q}^s) \), such that the one-parameter Noether group of transformation
\(\boldsymbol{\Phi}^{(\mathrm{N})}_\varepsilon : \mathbb{R}^+ \times \mathbb{R}^n \rightarrow \mathbb{R}^+ \times \mathbb{R}^n\), where
\begin{equation}\label{Eq.(3.67)}
(t,q^s) \mapsto (\bar{t},\bar{q}^s) = \boldsymbol{\Phi}^{(\mathrm{N})}_\varepsilon (t,q^s) = \left( \alpha (t,q^l;\varepsilon), \psi^s (t,q^l;\varepsilon) \right),
\end{equation}
with the infinitesimal transformation equations

\[t \rightarrow \bar{t} = a(t, q^i; \varepsilon) = t + \varepsilon \tau(t, q^i) + \mathcal{O}(\varepsilon^2),\]
\begin{equation}\label{Eq.(3.68)}
q^s \rightarrow \bar{q}^s = \psi^s(t, q^i; \varepsilon) = t + \varepsilon \xi^s(t, q^i) + \mathcal{O}(\varepsilon^2),
\end{equation}
holds invariant the functional action
\begin{equation}\label{Eq.(3.69)}
S[q^s] = \int_{t_1}^{t_2} L(t, q^i, \dot{q}^i) \mathrm{d}t,
\end{equation}
then, the Rand-Trautman identity can be written as \eqref{Eq.(3.65)}, where
\begin{equation}\label{Eq.(3.70)}
\mathbf{X}^{[1]} := \tau \frac{\partial}{\partial t} + \xi^k \frac{\partial}{\partial q^k} + (\dot{\xi}^k - \dot{\tau} \dot{q}^k) \frac{\partial}{\partial \dot{q}^k},
\end{equation}
is the first-order prolongation of the generator vector $\mathbf{X} = \tau \partial_t + \xi^k \partial_{q^k}$ associated to the Noether symmetry group \eqref{Eq.(3.67)}.

Here, for the purposes of calculations, we consider the gauge term to be zero: $G = 0$. The Rand-Trautman identity \eqref{Eq.(3.65)} then is written in a simpler form:
\begin{equation}\label{Eq.(3.71)}
\frac{\partial L}{\partial q} \xi + \frac{\partial L}{\partial \dot{q}} \dot{\xi} + \frac{\partial L}{\partial t} \tau - H\dot{\tau} = 0.
\end{equation}
Using the total derivative operator with respect to the cosmic time $t$, ${}^{\textbf{.}}$ = d/dt, we have the following relations for the infinitesimals $\tau(t, q)$ and $\xi(t, q)$ of the Noether generator: $\dot{\tau} = \tau_t + \dot{q}\tau_q$ and $\dot{\xi} = \xi_t + \dot{q}\xi_q$. Plugging these relations into the Rand-Trautman identity \eqref{Eq.(3.71)}, and then by simplifying it, we get the following relation
\begin{equation}\label{Eq.(3.72)}
q^{-2\alpha} [-2a \dot{q}^2 \xi + 2 \dot{q}(\xi_t + \dot{q}\xi_q) - (\dot{q}^2 - Cq^2)(\tau_t + \dot{q}\tau_q)] = 0.
\end{equation}
In general, $q^{-2\alpha} \neq 0$, so in order this relationship to always be established, it is necessary and sufficient that the expression inside the brackets is equal to zero
\begin{equation}\label{Eq.(3.73)}
Cq^2 \tau_t + (2\xi_t + Cq^2 \tau_q) \dot{q} + (2 \xi_q - \tau_t - 2a \xi) \dot{q}^2  -\tau_q \dot{q}^3 = 0.
\end{equation}
In order Eq. \eqref{Eq.(3.73)} to hold for all the values of the variables \(t\) and \(q\), it is necessary and sufficient that the coefficients of various powers of velocity \(\dot{q}\) are separately and independently equal to zero. Setting all coefficients of the different powers of \(\dot{q}\) to zero, yields to the following equations:
\begin{equation}\label{Eq3.74}
\ \ \ \ \ \ \ \ \ \ \ \ \ \ \ \ \ \partial_t\tau = 0,
\end{equation}
\begin{equation}\label{Eq3.75}
\ \ \ 2\partial_t\xi + Cq^{2}\partial_q\tau = 0,
\end{equation}
\begin{equation}\label{Eq3.76}
 2\partial_q\xi - \partial_t\tau - 2\alpha\xi = 0,
\end{equation}
\begin{equation}\label{Eq3.77}
\ \ \ \ \ \ \ \ \ \ \ \ \ \ \ \ \ \ \partial_q\tau = 0.
\end{equation}
Eqs. \eqref{Eq3.74}-\eqref{Eq3.77} form a system of first-order PDEs for the infinitesimal \(\tau(t,q)\) and \(\xi(t,q)\). These equations known as ``Killing equations''.
If one can solve this system of differential equations for the functions \(\tau(t,q)\) and \(\xi(t,q)\), simultaneously, then the generators of the Noether point symmetry will be obtained. In the next step, having the generators of the group and then applying the ``Lie's First Fundamental Theorem'' \cite{71}, one can achieve the Noether symmetry group itself, which keeps the functional action \eqref{Eq.(3.54)} invariant. Now, we solve the Killing equations \eqref{Eq3.74}-\eqref{Eq3.77}. Among them, Eqs. \eqref{Eq3.74} and \eqref{Eq3.77} means that \(\tau\) does not depend explicitly on the variables \(t\) and \(q\), so \(\tau\) is a constant, \(\tau=c_{1}\). Substituting Eqs. \eqref{Eq3.74} and \eqref{Eq3.77} into Eqs. \eqref{Eq3.75} and \eqref{Eq3.76} give us \(\xi_{t}=0\) and \(-\alpha\xi+\xi_{q}=0\).
The Killing equation \(\xi_{t}=0\) means that \(\xi\) does not depend explicitly on \(t\), so that \(\xi=\xi(q)\). Therefore, the partial derivative in the Killing equation \(-\alpha\xi+\xi_{q}=0\) may be replaced with total derivative, that is, \({\rm d}\xi/{\rm d}q-\alpha\xi=0\). Integrating this first-order ODE gives a solution as \(\xi=c_{2}{\rm e}^{\alpha q}\), where \(c_{2}\) is the integration constant. Thus, the general solutions of Eqs.
\eqref{Eq3.74}-\eqref{Eq3.77} become
\begin{equation}\label{Eq3.78}
\tau(t,q)=c_{1},\quad\xi(t,q)=c_{2}{\rm e}^{\alpha q}.
\end{equation}

Now, by using the functions \eqref{Eq3.78}, therefore, we have the generator of the one-parameter Noether group of point symmetry \eqref{Eq.(3.56)} as: \({\bf X}^{(\mathrm{N})}=c_{1}\partial_{t}+c_{2}{\rm e}^{\alpha q}\partial_{q}\). It can be seen that this vector is a linear combination of the vectors \(\partial_{t}\) and \({\rm e}^{\alpha q}\partial_{q}\). If these vectors are denoted respectively by \({\bf X}_{1}^{({\rm N})}\) and \({\bf X}_{2}^{({\rm N})}\), then, we see that the given Lagrangian function has two generator vectors: \({\bf X}_{1}^{({\rm N})}=\partial_{t}\), \({\bf X}_{2}^{({\rm N})}=e^{\alpha q}\,\partial_{q}\). The Noether point symmetry such that \({\bf X}_{1}^{({\rm N})}=\partial_{t}\) is its infinitesimal generator reads \({\bf\Phi}_{\varepsilon,\,1}^{({\rm N})}\dvtx\,{\mathbb{R}}^{+}\times\,{\mathbb{R}}\to{\mathbb{R}}^{+}\times\,{\mathbb{R}}\), where
\begin{equation}\label{Eq3.79}
(t,q)\mapsto(\bar{t},\bar{q})={\bf\Phi}_{\varepsilon,\,1}^{({\rm N})}(t,q)=\big{(}\alpha(t,q;{\varepsilon}),\psi(t,q;{\varepsilon})\big{)}=(t+{\varepsilon},q).
\end{equation}
The map \eqref{Eq3.79} is a time translation transformation. We are going to find a Noether point symmetry \({\bf\Phi}_{\varepsilon,\,2}^{({\rm N})}\) such that \({\bf X}_{2}^{({\rm N})}=e^{\alpha q}\,\partial_{q}\) is its infinitesimal generator. For this purpose, we use ``Lie's First Fundamental Theorem'' \cite{71}.
According to this theorem we have
\begin{equation}\label{Eq3.80}
\frac{{\rm d}}{{\rm d}\varepsilon}(\bar{t},\bar{q})={\bf X}_{2}^{({\rm N})}=(0,e^{\alpha\bar{q}}),
\end{equation}
with the initial condition \((\bar{t},\bar{q})|_{\varepsilon=0}=(t,q)\). Integrating of both sides of these differential equations and using the initial condition, one can easily obtain the transformation functions \(\alpha(t,q;\varepsilon)\) and \(\psi(t,q;\varepsilon)\) for the \({\bf\Phi}^{(\rm N)}_{\varepsilon,2}\) as follows:
\begin{equation}\label{Eq3.81}
\alpha(t,q;\varepsilon)=t,\quad\psi(t,q;\varepsilon)=-\frac{1}{\alpha}\ln( \mathrm{e}^{-\alpha q}-\alpha\varepsilon).
\end{equation}
Therefore, the Noether point symmetry associated to the generator \(\mathbf{X}^{(\rm N)}_{2}=\mathrm{e}^{\alpha q}\,\partial_{q}\) becomes: \({\bf\Phi}^{(\rm N)}_{\varepsilon,2}\colon\mathbb{R}^{+}\times\mathbb{R} \to\mathbb{R}^{+}\times\mathbb{R} \)
\begin{equation}\label{Eq3.82}
(t,q)\mapsto(\bar{t},\bar{q})={\bf\Phi}^{(\rm N)}_{\varepsilon,2}(t,q)=\big{(}t,-\alpha^{-1}\ln(\mathrm{e}^{-\alpha q}-\alpha\varepsilon)\big{)}.
\end{equation}

Now, according to Theorem 3.1 (Noether's Theorem) the Noether charge associated to the Noether point symmetry \eqref{Eq.(3.56)} is conserved
\begin{equation}\label{Eq3.83}
I(t,q,p):=G+H\tau-p\xi=\text{const.}
\end{equation}
Using the system of equations \eqref{Eq.(3.39)} for the DS \(S^{2}_{1}\), we have \(\dot{q}=\partial H/\partial p=pq^{-2\alpha}\). Hence, the generalized momentum is obtained to be \(p=\dot{q}q^{2\alpha}\). Substituting gauge term \(G=0\) and generalized momentum \(p=\dot{q}q^{2\alpha}\) into Eq. \eqref{Eq3.83} we then get
\begin{equation}\label{Eq3.84}
I(t,q,\dot{q}q^{2\alpha})=H(q,\dot{q}q^{2\alpha})\tau-\,\,\dot{q}q^{2\alpha}\xi =\text{const.}
\end{equation}
This conserved quantity is a function of the variables of the extended tangent bundle \(\mathbb{R}^{+}\times\mathrm{T}^{*}\mathbb{Q}\) which
we denote by \(I_{\rm N}(t,q,\dot{q}):=I(t,q,\dot{q}q^{2\alpha}\,)\). Therefore, the conserved Noether charge associated to the Noether point symmetry \({\bf\Phi}^{(\rm N)}_{\varepsilon}(t,q)\) becomes: \(I_{\rm N}(t,q,\dot{q})=E(q,\dot{q})\tau-\dot{q}q^{2\alpha}\xi=c_{\rm N}\), where \(E(q,\dot{q})=H(q,\dot{q}q^{2\alpha})\) is the energy function of the particle, and \(c_{_{\rm N}}\) is a constant. It should be noted that the energy function \(E\) has identically the same value as the Hamiltonian function \(H\), but they are functions of different variables. \(E\) is a function of \(q\), \(\dot{q}\) (and possibly \(t\)), while \(H\) must always be expressed as a function of \(q\), \(p\) (and possibly \(t\)). The difference in their functional behavior has led to the use of different symbols (\(E\) for the energy function and \(H\) for the Hamiltonian function) to denote these quantities, even though they have the same numerical values: \(E(q,\dot{q})=H(q,p)\).
The infinitesimals associated to the Noether point symmetry \eqref{Eq3.79} are \(\tau=1\) and \(\xi=0\).
Substituting these functions into the conserved quantity \eqref{Eq3.84} we can get
\begin{equation}\label{Eq3.85}
I_{\rm N,1}(q,\dot{q})=E(q,\dot{q})=c_{_{\rm N,1}}\,,
\end{equation}
where \( c_{_{\rm N,1}} \) is a constant. Therefore, the first integral associated to the Noether point symmetry \( {\bf\Phi}^{(\rm N)}_{\varepsilon,1}(t,q) = (t + \varepsilon, q) \) with the infinitesimal generator \( {\bf X}^{(\rm N)}_1 = \partial_t \) is the conservation of energy.
In this way, the conserved quantity associated to the Noether point symmetry \eqref{Eq3.82} with the infinitesimal generator \( {\bf X}^{(\rm N)}_2 = e^{\alpha q} \partial_q \) is
\begin{equation}\label{Eq3.86}
I_{\rm N,2}(q, \dot{q}) = \dot{q} q^{-2\alpha} e^{\alpha q} = c_{\rm N,2}.
\end{equation}

For the sake of clarity the results obtained are summarized in Table 1;
we display the Noether point symmetries corresponding to the DE \eqref{Eq.(2.9)} and the invariants associated to these symmetries
\\
\begin{center}
		\small {{{\bf Table 1.}~ The Noether point symmetries of the DE \eqref{Eq.(2.9)} and \\$~~~~~~~~~~~~~~$the first integrals associated to these symmetries.}}\\
		{\small
			\renewcommand{\arraystretch}{1.5}{
\begin{tabular}{ l l  } \hline \hline
      Symmetries & Invariants (First integrals) \\ \hline
$ {\bf X}^{(\rm N)}_1 = \partial_t  $  &  $ I_{_{\rm N,1}}(q, \dot{q}) = \frac{1}{2} \dot{q}^2 q^{-2\alpha} - \frac{1}{2} C q^{-2\alpha+2} = E = c_{\rm N,1} $ \\

     ${\bf X}^{(\rm N)}_2 = e^{\alpha q} \partial_q$ & $I_{_{\rm N,2}}(q, \dot{q}) = \dot{q} q^{-2\alpha} e^{\alpha q} = c_{_{\rm N,2}}$ \\ \hline \hline

		\end{tabular}}}
\end{center}
$\\$
The general solution of the EFDE \eqref{Eq.(2.9)}  cannot be obtained by simultaneously solving Noether's invariants \eqref{Eq3.85} and \eqref{Eq3.86}, because these invariants are not independent from each other. Therefore, it seems that the Noether symmetry approach, although it provides us with useful information about the Hamiltonian structure of the cosmological model, is not suitable for solving DEs like our cosmological model. As we will show in Section \ref{section 4}, this idea is not correct so that one can use the HJE to find two independent invariants that are necessary to solve Eq. \eqref{Eq.(2.9)} and thus complete the solution of the problem.

\section{The IHJM to solve the DE of FLRW}\label{section 4}

Let us re-consider the FLRW cosmological model of Section \ref{section 2}.
We studied this model in Section \ref{section 3} via both Hojman formalism and Noether symmetry approach. As we mentioned earlier, in the Hojman formalism whose dynamical points are labeled by the 2-tuple \( x^i := (q, \dot{q}) \), and the force acting on the particle by the 2-tuple \( \Phi^i := \left( \dot{q}, F(q, \dot{q}) \right) \), the DE \eqref{Eq.(2.9)} can be rewritten as \eqref{3.1}. These equations form a set of the first-order ODEs on the tangent bundle \( {\rm T}\mathbb{Q} \) of the configuration space \( \mathbb{Q} = (q) \). In this formalism, the tangent bundle \( {\rm T}\mathbb{Q} \) consists of the configuration space and a set of tangent spaces \( {\rm T}_q\mathbb{Q} \), each attached to a point \( q \in \mathbb{Q} \). The Hojman's equations set the \( \dot{x}^i\)'s, \(i = 1, 2 \), the components of a vector field \( {\bf X} = x^i \partial_{x^i} \) in \( \mathfrak{X}({\rm T}\mathbb{Q}) \) equal to the components of the vector field \( {\bf \Phi} = \Phi^i \partial_{x^i} \in \mathfrak{X}({\rm T}\mathbb{Q}) \). Therefore, the EFDE of the FLRW model behaves like as a Hojman's DS where the governing equations of the particle in tangent bundle \( {\rm T}\mathbb{Q} \) are given by the system of two first-order ODEs \eqref{3.2}. In this formalism, by using the time independent invariant \eqref{Eq.(3.11)}, which is the energy of the DS associated to the infinitesimal generator, \( {\bf X} = \partial_t \) as one of both the Lie and Noether point symmetries of the DS, we can obtain the Lagrangian function \eqref{Eq.(3.24)} and then the Hamiltonian function \eqref{Eq.(3.42)}.

Now, we apply the method introduced in Appendix B for the FLRW cosmological model with the Hamiltonian function \eqref{Eq.(3.42)}.
 Substituting this function in the HJE \eqref{Eq4.14}, we arrive at the following equation
\begin{equation}\label{Eq4.23}
\frac{1}{2}\Big{(}\frac{\partial S}{\partial q}\Big{)}^{2}q^{2\alpha}-\frac{1}{2}Cq^{2-2\alpha}+\frac{\partial S}{\partial t}=0.
\end{equation}
In order to solve the PDE \eqref{Eq4.23}, we separate the Hamilton's principal function in the form \(S(t,q,P)=S_{1}(q,P)+S_{2}(t,P)\).
Then we have \(\partial S/\partial q=\mathrm{d}S_{1}/\mathrm{d}q\) and \(\partial S/\partial t=\mathrm{d}S_{2}/\mathrm{d}t\).
By putting these equations into Eq. \eqref{Eq4.23}, we arrive at the following first-order ODE
\begin{equation}\label{Eq4.24}
\frac{1}{2}\Big{(}\frac{\mathrm{d}S_{1}}{\mathrm{d}q}\Big{)}^{2}q^{2\alpha}-\frac{1}{2}Cq^{2-2\alpha}=-\frac{\mathrm{d}S_{2}}{\mathrm{d}t},
\end{equation}
As it can be seen both sides of above equation must be equal to a constant value $P$, which is called the ``separation constant''.
Therefore, we have two first-order ODEs
\begin{eqnarray}
\frac{1}{2}\Big{(}\frac{\mathrm{d}S_{1}}{\mathrm{d}q}\Big{)}^{2}\,q^{2\alpha}-\frac{1}{2}\,Cq^{2-2\alpha}&=&P,\label{Eq4.25}\\
\frac{\mathrm{d}S_{2}}{\mathrm{d}t}&=&-P.\label{Eq4.26}
\end{eqnarray}
Integrating from Eqs. \eqref{Eq4.25} and \eqref{Eq4.26} we obtain the following solutions
\begin{eqnarray}
S_{1}(q,P)&=&\int q^{-\alpha}\sqrt{Cq^{2-2\alpha}+2P}\,\mathrm{d}q,\label{Eq4.27} \\
S_{2}(t,P)&=&-Pt.\label{Eq4.28}
\end{eqnarray}
Then, one can obtain the Hamilton's principal function as
\begin{eqnarray}\label{Eq4.29}
S(t,q,P)&=&S_{1}(q,P)+S_{2}(t,P)\nonumber \\
         &=&-Pt+\int q^{-\alpha}\sqrt{Cq^{2-2\alpha}+2P}\,\mathrm{d}q.
\end{eqnarray}
It can be shown that the function \eqref{Eq4.29} satisfies to the Hessian condition $\partial^{2}S/\partial q\partial P\neq 0$. So, the generating function \eqref{Eq4.29} is the complete solution of the HJE \eqref{Eq4.23}. Now, inserting this generating function into the first equation of the system of Eqs. \eqref{Eq4.15} we can find gives the following equation
\begin{equation}\label{Eq4.30}
Q=-t+\int\frac{q^{-\alpha}\mathrm{d}q}{\sqrt{Cq^{2-2\alpha}+2P}}\,.
\end{equation}
To calculate the integral in Eq. \eqref{Eq4.30}, one may define the new variable: \( u := q^{-\alpha+1} \). We calculate this integral in three different cases \( C > 0, \, C = 0\) and  \(C < 0 \), separately.
Finally, we can get
\begin{equation}\label{Eq4.36}
Q =
\begin{cases}
-t + \frac{1}{\sqrt{|\beta(1 - \alpha)|}} \sin^{-1} \left( \frac{\sqrt{|C|}}{\sqrt{2P}} q^{1-\alpha} \right) & C < 0, \\
-t + \frac{q^{1-\alpha}}{\sqrt{2P}(1 - \alpha)} & C = 0, \\
-t + \frac{1}{\sqrt{\beta(1 - \alpha)}} \sinh^{-1} \left( \frac{\sqrt{C}}{\sqrt{2P}} q^{1-\alpha} \right) & C > 0,
\end{cases}
\end{equation}
where the constants $\alpha$ and $\beta$ are defined according to Eqs. \eqref{Eq.(2.8)}.
In the following, inserting the generating function \eqref{Eq4.29} into the second equation of the system of Eqs. \eqref{Eq4.15} we can get
\begin{equation}\label{Eq4.37}
p = q^{-\alpha} \sqrt{C q^{2-2\alpha} + 2P}.
\end{equation}
Now, solving the system of equations \eqref{Eq4.36} and \eqref{Eq4.37} simultaneously give us the canonical variables $Q$ and $P$ in terms of the variables $p$, $q$ and $t$.
Again, we perform this calculation in three different cases: $C<0$, $C=0$ and $C>0$, separately:

(a) Case $C>0$. In this case, by solving the algebraic equations \eqref{Eq4.36} and \eqref{Eq4.37} for the variables $P$ and $Q$ we find that
\begin{eqnarray}
P&=&\frac{1}{2}p^{2}q^{2\alpha}-\frac{1}{2}Cq^{2-2\alpha},\label{Eq4.38}\\
Q&=&-t+\frac{1}{\sqrt{\beta(1-\alpha)}}\sinh^{-1}\Big(\frac{\sqrt{C}q^{-\alpha+1}}{\sqrt{p^{2}q^{2\alpha}-Cq^{-2\alpha+2}}}\Big).\label{Eq4.39}
\end{eqnarray}
We note that Eqs. \eqref{Eq4.38} and \eqref{Eq4.39} together specify the CT in the case where the cosmological constant $\Lambda$ is positive.
Indeed, this transformation keeps the HCEs invariant.

(b) Case $C=0$. In the same way, it can be shown that the equations of the CT in the case $C=0$ are
\begin{eqnarray}
P&=&\frac{1}{2}p^{2}q^{2\alpha},\label{Eq4.40}\\
Q&=&-t+\frac{q^{1-2\alpha}}{(1-\alpha)p}.\label{Eq4.41}
\end{eqnarray}

(c) Case $C<0$. In this case, the equations of the CT become
\begin{eqnarray}
P&=&\frac{1}{2}p^{2}q^{2\alpha}+\frac{1}{2}|C|q^{2-2\alpha},\label{Eq4.42}\\
Q&=&-t+\frac{1}{\sqrt{|\beta(1-\alpha)|}}\sin^{-1}\left(\frac{\sqrt{|C|q^{-\alpha+1}}}{\sqrt{p^{2}q^{2\alpha}+|C|q^{-2\alpha+2}}}\right).\label{Eq4.43}
\end{eqnarray}
According to the definition of the generalized momentum conjugate to the generalized coordinate $q$, Eq. \eqref{Eq.(3.36)}, we have $p=\dot{q}q^{-2\alpha}$.

Now we are in position to extract the invariants associated to the EFDE \eqref{Eq.(2.9)} from the CT.
In this case, by inserting $p=\dot{q}q^{-2\alpha}$ into the CTs \eqref{Eq4.38}-\eqref{Eq4.43} one can get the corresponding invariants.
The results obtained are summarized in Table 2;
we display the invariants corresponding to the three different cases \( C > 0, \, C = 0, \, C < 0 \).
Note that in Table 2, the $P$ and $Q$ on the right side of the invariants are constants.
\\
\begin{center}
		\small {{{\bf Table 2.}~ The invariants associated to the EFDE \eqref{Eq.(2.9)} from the CT.}}\\
		{\small
			\renewcommand{\arraystretch}{1.5}{
\begin{tabular}{ l l  } \hline \hline
      Different cases & Invariants (First integrals)  \\ \hline
$C<0 ~(\Lambda<0)$  &  $ I_{\rm Hj,1}(q,\dot{q})=\frac{1}{2}\,\dot{q}^{2}q^{3w+1}-\frac {\Lambda}{6}\,q^{3(w+1)}=P, $ \\
\vspace{1mm}
       & $ I_{\rm Hj,2}(t,q,\dot{q})=-\frac{3(w+1)}{2}t+\frac{1}{\sqrt{-\Lambda/3}}\tan^{-1}\Big{(}\frac{q}{\dot{q}}\sqrt{-\Lambda/3}\Big{)}=(1-\alpha)Q $ \\

$C=0~ (\Lambda=0)$  &  $ I_{\rm Hj,1}(q,\dot{q})=\frac{1}{2}\,\dot{q}^{2}q^{3w+1}=P,$ \\
\vspace{1mm}
       & $ I_{\rm Hj,2}(t,q,\dot{q})=\frac{-3(w+1)}{2}t+\frac{q}{\dot{q}}=(1-\alpha)Q.$ \\

$C>0 ~(\Lambda>0)$  &  $ I_{\rm Hj,1}(q,\dot{q})=\frac{1}{2}\,\dot{q}^{2}q^{3w+1}-\frac {\Lambda}{6}\,q^{3(w+1)}=P, $ \\
       & $I_{\rm Hj,2}(t,q,\dot{q})= -\frac{3(w+1)}{2}t+\frac{1}{\sqrt{\Lambda/3}}\tanh^{-1}\Big{(}\frac{q}{\dot{q}}\sqrt{\Lambda/3}\Big{)}=(1-\alpha)Q. $ \\ \hline \hline

		\end{tabular}}}
\end{center}
$\\$
It is worth noting that the invariants obtained by the CT in Table 2 are the same invariants that we previously obtained by the ISGM \cite{1,84}.
Solving these as a system of two algebraic equations with two unknowns $q$ and $\dot{q}$ will be easy to complete the problem.
Therefore, solving the systems of equations formed by the invariants from Table 2 separately gives the variable $q$ in terms of time $t$ and constants $P$ and $Q$ as follows
\begin{equation}\label{Eq4.61}
q(t) =
\begin{cases}
\Big{(}-\frac{6P}{\Lambda}\Big{)}^{\frac{1}{3(w+1)}}\sin^{\frac{2}{3(w+1)}}\Big[\frac{w+1}{2}\sqrt{-3\Lambda}(t+Q)\Big], & \Lambda<0, \\
\big(3P(w+1)\big)^{\frac{1}{3(w+1)}}(t+Q)^{\frac{2}{3(w+1)}}, & \Lambda=0, \\
\Big{(}\frac{6P}{\Lambda}\Big{)}^{\frac{1}{3(w+1)}}\sinh^{\frac{2}{3(w+1)}}\Big[\frac{w+1}{2}\sqrt{3\Lambda}(t+Q)\Big], & \Lambda>0,
\end{cases}
\end{equation}
which is the general solution of the EFDE \eqref{Eq.(2.9)} obtained from solving the HJE \eqref{Eq4.23} by using the Hamilton-Jacobi invariants.
As we have seen, this solution was obtained by solving the HJE \eqref{Eq4.23} and using the CT ${\bf\Phi}_{\rm CT}(t,q,p)\mapsto(t,Q,P)$ that keeps the HCEs invariant.
In Hamiltonian dynamics, since we deal with CT-symmetry, the independent invariants of the DS can be obtained when the desired cosmological model can be described as a DS. If there is a time-independent constant of motion, then the Hamiltonian formalism of the DS is a suitable way to study the cosmological model, because for the DS, Lagrangian and Hamiltonian functions can be constructed by the Hojman's method.
Therefore, by using the HJE one can find the CT that keeps the HCEs invariant.
When we achieve this transformation, we have actually managed to find two independent invariants $\big{(}I_{\rm HJ,1},I_{\rm HJ,2}\big{)}=\big{(}(1-\alpha)Q,P\big{)}$.
In this way, the desired problem can be solved by the IHJM.
This solution method has another advantage, and this is that in addition to the complete solution of the problem, the IPCV is also solved.
The construction of the Lagrangian and Hamiltonian functions for a cosmological model is of particular importance,
because by doing this, by constructing the Lagrangian function (and thus the Hamiltonian function),
it is possible to construct a Hamiltonian structure with Poisson and Lagrange Brackets using Hojman's method for the DS.
For example, if it is necessary to raise quantum arguments in the cosmological model, the IHJM is also used for those cosmological models in which we have the equations of the field theory and look for its Lagrangian function, can be used.

\section{Conclusion}\label{section 5}

In this study, we presented a new method for solving the DEs of the FLRW cosmological model spatially flat ($k=0$) in the presence of the cosmological constant and within the framework of the theory of GR.
We considered the universe as a perfect fluid with the equation of state $p=w\rho$, $-1\leq w\leq 1$. This method, which we called in this study the IHJM, is analytical and systematic. It can be used, in addition to solving the DEs of the FLRW cosmological model, for any cosmological model whose dynamics behaves like a single particle so that the Hamiltonian formalism can be used.
The IHJM is algorithmic and consists of several steps:
in the first step, we showed that the dynamics of the FLRW cosmological model in the presence of $\Lambda$ and within the framework of the GR behaves like a single-particle dynamic
in a mini-super one-dimensional space ${\mathbb{Q}}=(a)$ called the configuration space.
This correspondence between the cosmological model and the mini-supper configuration space ${\mathbb{Q}}=(a)$ helped us to use the dynamics of the single-particle in the space ${\mathbb{Q}}=(a)$
to solve the DEs of the FLRW cosmological model.
In the next stage of our research, we showed that in the configuration space, the energy of the particle is one of the constants of motion of the DS.
Using this constant of motion and via the Hojman's method \cite{84}, we obtained the Lagrangian and Hamiltonian functions of the particle in the tangent bundle space $\rm T{\mathbb{Q}}$,
and the cotangent bundle space $\rm T^{*}{\mathbb{Q}}$, respectively.
Then, we presented a Hamiltonian structure with the Poisson Brackets matrix $J^{ij}$ and the Lagrange Brackets matrix $\sigma_{ij}$ in the Hamiltonian formalism.
In the continuation of the discussion, we first presented a concise review of the Noether point symmetry, the Noether theorem and the Rand-Trautman identities, then
by using the Killing equations we obtained the Noether symmetries of the particle in the configuration space.
We have shown that these symmetries are not sufficient to solve the DE of the particle,
so we must turn to the Hamilton-Jacobi theorem (Theorem B.1).
Then, having the Hamiltonian function and using the Hamilton-Jacobi theorem, we found a CT that kept the HCEs invariant.
By this transformation, we were able to explicitly obtain two independent invariants (or first integrals) $I_{\rm HJ,1}(a,\dot{a})=P\;$and $I_{\rm HJ,2}(t,a,\dot{a})=(1-\alpha)Q$. These invariants were exactly the same ones that we have previously achieved in other works using the ISGM \cite{1,72}.
In fact, extracting these two independent invariants via the IHJM for the DE of the particle in the mini-supper space is not only an exact analytical solution of the DEs of our cosmological model, but also a solution of the IPCV. In other words, by using the IHJM, in addition to obtaining the general solution of the DEs of the desired cosmological model, another important result can be extracted from the DE of the particle in the configuration space.
This result is extracting the Lagrangian function of the DS from the DE of the particle,
i.e. solving the IPCV without resorting to the Helmholtz conditions
and whether the necessary conditions for the existence of the Lagrangian function hold or not.
Therefore, caring out this part of the research, i.e., obtaining the Lagrangian function for the DEs of a given cosmological model in theories such as the Rastall theory of gravity \cite{85}, in which having the field equations one seeks to find its Lagrangian function, it can be of particular importance.
According to the same results obtained through the IHJM and ISGM for the cosmological FLRW model,
there seems to be a close connection between the relationships of these two solution methods.
Perhaps combining these two solution methods can provide us with a comprehensive integration theory for solving DEs in gravity and cosmology.

\appendix

\section{\label{app.A} Some of the definitions and theorems used throughout the text: Poisson and Lagrange brackets matrices, Hojman's formula, etc.}

In this Appendix, we review some of the definitions and theorems used throughout the text of the paper.
\\
\\
\textbf{Definition A.1} (Poisson Brackets Matrix). \emph{For the} \emph{FLRW} \emph{cosmological model, whose dynamic is imagined as the motion of a particle in two-dimensional tangent bundle} \(\mathrm{T}\mathbb{Q}\) \emph{with the local coordinates} \(x^{k}:=(q,\dot{q})\), \emph{and} \(I\big{(}x^{k}\big{)}\) \emph{is a time-independent constant of motion for this particle, the antisymmetric matrix} \(J^{ij}\), \(i,j=1,2\), \emph{that holds under the following conditions, which is called the Poisson Brackets matrix} \cite{2},

(a) \emph{Antisymmetric condition}:
\begin{equation}\label{Eq.(3.13)}
J^{ij}=-J^{ji},
\end{equation}
\(\,\,\,\,\,\,\,\,\, \)(b) \emph{the Jacobi identity}:
\begin{equation}\label{Eq.(3.14)}
J^{ij}{}_{,k}J^{kl}+J^{jl}{}_{,k}J^{ki}+J^{li}{}_{,k}J^{kj}=0,
\end{equation}
\(\,\,\,\,\,\,\,\,\, \)(c) \emph{and Hamilton equations}:
\begin{equation}\label{Eq.(3.15)}
\Phi^{i}\big{(}x^{k}\big{)}=J^{ij}\,\frac{\partial H}{\partial x^{j}}:=\big {[}x^{i},H\big{]},
\end{equation}
\emph{where \(\Phi^{i}\)'s, \(i=1,2\), are the components of the force acting on the particle in the} \emph{DS} \emph{\(S_{2}^{1}\) and \(H=I\big{(}x^{k}\big{)}\) is its Hamiltonian function. Also, \([A,B]\) is called Poisson Brackets which is defined for any pair of dynamical variables \(A(x^{i})\) and \(B(x^{j})\) as follows:}
\begin{equation}\label{Eq.(3.16)}
[A,B]:=\frac{\partial A}{\partial x^{i}}J^{ij}\frac{\partial B}{\partial x^{j}},
\end{equation}
\emph{where Einstein summation convention is used. Notice that in the Jacobi identity} \eqref{Eq.(3.14)}, \emph{the symbol ``,'' denotes the partial derivative with respect to that variable. For example, \(J^{ij}{}_{,k}=\partial J^{ij}/\partial x^{k}\).}

Note that substituting \(A=x^{i}\) and \(B=H\) into Eq. \eqref{Eq.(3.16)} leads to \eqref{Eq.(3.15)}:

\[\big{[} x^{i},H\big{]}=\frac{\partial x^{i}}{\partial x^{k}}J^{kj}\frac{\partial H}{\partial x^{j}}=\delta^{i}{}_{k}J^{kj}\frac{\partial H}{\partial x^{j}}=J^{ij}\frac{\partial H}{\partial x^{j}}\,,\]
where \(\delta^{i}{}_{k}\) is the Kronecker delta: \(\delta^{i}{}_{k}=0\) for \(i\neq k\), and \(\delta^{i}{}_{k}=1\,\) for \(i=k\).

\(\)
\\
\textbf{Theorem A.1} (Hojman's Formula, 2014). \emph{For a particle in the} \emph{DS} \(S^{1}_{2}\) \emph{with time-independent constant of motion} \(I(x^{k})\) \emph{and the Poisson Brackets matrix} \(J^{ij}\big{(}x^{k}\big{)}\), \(i,j=1,2\), \emph{the Lagrangian function} \(L\big{(}x^{k}\big{)}\) \emph{is defined as follows} \cite{2}:
\begin{equation}\label{Eq.(3.17)}
L(x^{k})=l_{1}\big{(}x^{k}\big{)}\dot{x}^{1}-I(x^{k}),
\end{equation}
\emph{where the function \(l_{1}\big{(}x^{k}\big{)}\) is the general solution of the} \emph{PDE}:
\begin{equation}\label{Eq.(3.18)}
\frac{\partial l_{1}\big{(}x^{k}\big{)}}{\partial x^{2}}=\frac{1}{J^{12}(x^{k})}.
\end{equation}
\emph{In fact, the Lagrangian function} \(L\) \emph{of the particle, is a map from the tangent bundle} \(\mathrm{T}\mathbb{Q}\) \emph{into the set of real numbers} \(\mathbb{R}\), \emph{that is}: \(L:\ \mathrm{T}\mathbb{Q}\rightarrow\mathbb{R}\), where Eq. \eqref{Eq.(3.17)} holds. \emph{Therefore, \(L\) is a real-valued function on the tangent bundle of the configuration space: \(L\in\mathcal{F}(\mathbb{T}\mathbb{Q})\).}

\(\)
\\
\textbf{Proof.} See \cite{2}.

\(\)

It should be noted that Eq. \eqref{Eq.(3.17)} for calculating the lagrangian function, which is attributed to S. A. Hojman \cite{2}, then, we have named in this study ``Hojman's formula'' in honor of its author.

\(\)
\\
\textbf{Definition A.2} (Lagrange Brackets Matrix).
\emph{For a particle in the} \emph{DS} $S_2^1$ \emph{with a Poisson Brackets matrix} $J^{ij}(x^k)$, $i,j = 1,2$, \emph{the Lagrange Brackets matrix} $\sigma_{ij}, i,j = 1,2$, \emph{is defined as follows} \cite{2}:
\begin{equation}\label{Eq.(3.19)}
\sigma_{ij} = -J_{ij} \cdot
\end{equation}
\emph{where} $J_{ij}$ \emph{is the inverse of the Poisson Brackets matrix} $J^{ij}$, \emph{that is}, $J^{ij}J_{jl} = \delta^i_{~l}$, \emph{and} $x^k := (x^1,x^2) := (q,\dot{q})$.

\(\)

It can be easily shown that the Poisson Brackets matrix $J^{lj}$ and the Lagrange Brackets matrix $\sigma_{ij}$ are satisfy into the relationship: $J^{li}\sigma_{ij}=-\delta^l_{~j}.$ To do this,
we multiply both sides of Eq. \eqref{Eq.(3.19)} by $J^{li}$ and then summing on the dummy index $i$, and finally using the Einstein summation convention. The result of these calculations will be the desired equation.

The above discussion can be summarized as follows: when a time-independent invariant $I(x^k)$ is known, then the Hamiltonian structure can be defined by choosing the invariant $I(x^k)$ as the Hamiltonian function $H = I(x^k)$ and therefore, matrix $J^{ij}(x^k)$ is completely determined by the conditions \eqref{Eq.(3.13)}, \eqref{Eq.(3.14)} and \eqref{Eq.(3.15)}. Once the Poisson Brackets matrix $J^{ij}$ is determined, then by using the PDE \eqref{Eq.(3.18)}, one can find the function $l_1(x^k)$. Finally, having function $l_1(x^k)$, the Hojman's formula \eqref{Eq.(3.17)} enables us to construct the Lagrangian function $L(x^k)$ for the particle in the \(\mathrm{DS}\) $S_2^1$.
\\
\\
\textbf{Definition A.3} (Legendre Transformation). \emph{Let us consider a DS with} \( n \) \emph{degrees of freedom and with configuration space} \(\mathbb{Q} = q^{\alpha} := (q^1, \dots, q^n)\). \emph{Let} \( L(q^{\alpha}, \dot{q}^{\alpha}) \) \emph{be the Lagrangian function of this DS which is defined on the tangent bundle} \( \mathrm{T}\mathbb{Q} \). \emph{The Legendre transformation of the Lagrangian function} \( L \) \emph{is given by the fibre derivative of the function \( L \) with respect to the fibres} \( q^{\alpha,} \)\emph{s, which is defined by the map} \(\mathrm{\textbf{FL}}: \mathrm{T}\mathbb{Q} \rightarrow \mathrm{T}^{*}\mathbb{Q}\), where
\begin{equation}\label{Eq.(3.31)}
(q^\alpha, \dot{q}^\alpha) \mapsto \mathrm{\textbf{FL}}(q^\alpha, \dot{q}^\alpha) = (q^\alpha, p_\alpha),
\end{equation}
\emph{in which} \( p_\alpha := \partial L/\partial \dot{q}^\alpha \in \mathrm{T}_{q^\alpha}^{*}\mathbb{Q}\text{'s}, \alpha = 1, \dots, n \). \emph{As we have seen already, the generalized velocities} \( \dot{q}^\alpha \in \mathrm{T}_{q^\alpha}\mathbb{Q}\text{'s} \), \emph{at the points} \( q^\alpha \in \mathbb{Q} \) \emph{are called its fibres of the tangent bundle} \( \mathrm{T}\mathbb{Q} \). \emph{The quantities} \( p_\alpha \in \mathrm{T}_{q^\alpha}^{*}\mathbb{Q}\text{'s} \), \emph{is usually called the canonical momentum conjugate to the configuration variable} \( q^\alpha\text{'s}, \alpha = 1, \dots, n \). \emph{Notice, in some papers, for example} \cite{52,53,54,55,56,57}, \emph{the map}, \( \mathrm{\textbf{FL}}: \mathrm{T}\mathbb{Q} \rightarrow \mathrm{T}^{*}\mathbb{Q} \), \emph{where} \( (q^\alpha, \dot{q}^\alpha) \mapsto (q^\alpha, p_\alpha), p_\alpha = \partial L/\partial \dot{q}^\alpha \), \emph{is known as the Legendre transformation, and sometimes is said the fibre derivative of the Lagrangian function.}

\( \)

To transition from the Lagrangian formulation to the Hamiltonian formulation and vice versa, the map \eqref{Eq.(3.31)} acts as a bridge between the tangent bundle \(\mathbb{T}\mathbb{Q}\) and the cotangent bundle \(\mathbb{T}^{*}\mathbb{Q}\).

\(\)
\\
\textbf{Definition A.4} (Non-degenerate Lagrangian System). \emph{A Lagrangian function \(L(q^{s},\dot{q}^{s})\) of the DS \(S_{n}^{2}\), is said to be non-degenerate if the \(n\times n\) Hessian matrix of the Lagrangian function \(L(q^{s},\dot{q}^{s})\), that is,}
\begin{equation}\label{Eq.(3.32)}
M_{\mathrm{L}}:=\omega_{\alpha\beta}(q^{s},\dot{q}^{s}):=\frac{\partial^{2}L}{ \partial\dot{q}^{\alpha}\partial\dot{q}^{\beta}}=\begin{pmatrix}\frac{\partial ^{2}L}{\partial\dot{q}^{1}\partial\dot{q}^{1}}&\cdots&\frac{\partial^{2}L}{ \partial\dot{q}^{1}\partial\dot{q}^{n}}\\ \vdots&\ddots&\vdots\\ \frac{\partial^{2}L}{\partial\dot{q}^{n}\partial\dot{q}^{1}}&\cdots&\frac{ \partial^{2}L}{\partial\dot{q}^{n}\partial\dot{q}^{n}}\end{pmatrix},
\end{equation}
\emph{be invertible everywhere on the tangent bundle} \(\mathrm{T}\mathbb{Q}\). \emph{In other words, the Hessian matrix} \(\omega_{\alpha\beta}\) \emph{be nonsingular, i.e., satisfies the Jacobi condition}:
\begin{equation}\label{Eq.(3.33)}
\det M_{\mathrm{L}}=\frac{\partial(p_{1},...,p_{n})}{\partial(\dot{q}^{1}, \cdots,\dot{q}^{n})}\neq 0.
\end{equation}
\emph{Such Lagrangian systems satisfies in the Jacobi condition \eqref{Eq.(3.33)}, are also said to be as hyperregular} \cite{55,56,57,58,59}.

\(\)
\\
\textbf{Theorem A.2} (Hamiltonian Function). \emph{Suppose that} \(L(q^{s},\dot{q}^{s})\) \emph{be the Lagrangian function for} \(S_{n}^{2}\) \emph{with the configuration space} \(\mathbb{Q}=q^{s}:=(q^{1},\cdots,q^{n})\), \emph{if the Lagrangian function} \(L(q^{s},\dot{q}^{s})\) \emph{is hyper-regular, then the Hamiltonian function} \(H(q^{s},p_{s})\) \emph{for the DS, exists and obtained by the map} \(\mathbf{\Phi}_{\mathrm{L}}\colon \mathcal{F}(\mathrm{T}\mathbb{Q})\to\mathcal{F}(\mathrm{T}^*\mathbb{Q})\), where
\begin{equation}\label{Eq.(3.34)}
L\mapsto\mathbf{\Phi}_{\rm L}(L)=H(q^s,p_s)=p_\alpha\dot{q}^\alpha(q^s,p_s)-L\big(q^\alpha,\dot{q}^\alpha(q^s,p_s)\big),
\end{equation}
\emph{in which} \(\dot{q}^{\alpha}(q^{s},p_{s})\)\emph{'s}, \(\alpha=1,\cdots,n\), \emph{are solutions of the system of equations} \(p_{\alpha}:=\partial L/\partial\dot{q}^{\alpha}\)\emph{'s, for the fibres of the tangent bundle,} \(\dot{q}^{\alpha}\)\emph{'s in terms of the canonical variables} \(q^{\alpha}\)\emph{'s, and} \(p_{\alpha}\)\emph{'s. The map} \eqref{Eq.(3.34)} \emph{that transforms the Lagrangian function into the Hamiltonian function is called the Legendre map of the Lagrangian function \(L\) and the equations}:

\( \)

\begin{equation}\label{Eq.(3.35)}
H(q^{s},p_{s}) =p_{\alpha}\dot{q}^{\alpha}(q^{s},p_{s})-L\big{(}q^{\alpha},\dot{ q}^{\alpha}(q^{s},p_{s})\big{)},
\end{equation}
\begin{equation}\label{Eq.(3.36)}
p_{\alpha} =\frac{\partial L(q^{s},\dot{q}^{s})}{\partial\dot{q}^{\alpha}}, \quad\alpha=1,\cdots,n,
\end{equation}
\emph{are said the Legendre transformations of the Lagrangian function.}

\(\)

The map \eqref{Eq.(3.34)}, which transforms the Lagrangian function \( L(q^s, \dot{q}^s) \) to the Hamiltonian function \( H(q^s, p_s) \), is actually the Legendre map \eqref{Eq.(3.31)}, but with the difference that map \eqref{Eq.(3.31)} transforms the tangent bundle space \( \mathrm{T}\mathbb{Q} \) to the cotangent bundle space \( \mathrm{T}^{*}\mathbb{Q} \), while by the map \eqref{Eq.(3.34)}, the Lagrangian function \( L(q^s, \dot{q}^s) \) transforms to the Hamiltonian function \( H(q^s, p_s) \). Hence, the map \eqref{Eq.(3.34)} is also called the ``Legendre map''. It should be noted that the system of the equations \eqref{Eq.(3.36)}, can be solved for the fibres \( \dot{q}^\beta \text{'s} \), in terms of the canonical variables \( q^\alpha \text{'s} \) and \( p_\alpha \text{'s} \), simultaneously, because it is assumed that the given Lagrangian function \( L(q^s, \dot{q}^s) \) is non-degenerate. Thus, the Legendre transformation maps the Lagrangian function into the Hamiltonian function. In the same way, by applying the inverse of the Legendre transformation, i.e., \( \mathbf{\Phi}_{\mathrm{L}}^{-1} \), the Hamiltonian function \( H(q^s, p_s) \) transforms into the Lagrangian function \( L(q^s, \dot{q}^s) \).

\(\)
\\
\textbf{Theorem A.3} (Inverse of the Theorem A.2). \emph{Suppose that the Hamiltonian function} \( H(q^s, p_s) \) \emph{is given for the DS} \( S_n^2 \), \emph{and the Jacobi condition}
\begin{equation}\label{Eq.(3.37)}
\det M_\mathrm{H} = \frac{\partial (\dot{q}^1, \cdots, \dot{q}^n)}{\partial (p_1, \cdots, p_n)} \neq 0,
\end{equation}
\emph{where}, \emph{the} \( n \times n \) \emph{matrix}

\[M_\mathrm{H} := \Omega^{\alpha \beta}(q^s, p_s) = \frac{\partial^2 H}{\partial p_\alpha \partial p_\beta} =
\begin{pmatrix}
\frac{\partial^2 H}{\partial p_1 \partial p_1} & \cdots & \frac{\partial^2 H}{\partial p_1 \partial p_n} \\
\vdots & \ddots & \vdots \\
\frac{\partial^2 H}{\partial p_n \partial p_1} & \cdots & \frac{\partial^2 H}{\partial p_n \partial p_n}
\end{pmatrix}\]
\emph{as the Hessian matrix of the Hamiltonian function} \( H(q^s, p_s) \), \emph{is fulfilled everywhere on the cotangent bundle} \( \mathrm{T}^{*}\mathbb{Q} \), \emph{then the Lagrangian function} \( L(q^s, \dot{q}^s) \) \emph{for the DS} \( S_n^2 \) \emph{exists, and obtained by the map} \(\mathbf{\Phi}_{\mathrm{L}}^{-1}: \mathcal{F}(\mathrm{T}^{*}\mathbb{Q}) \rightarrow \mathcal{F}(\mathrm{T}\mathbb{Q})\), \emph{where}
\begin{equation}\label{Eq.(3.38)}
H \mapsto \mathbf{\Phi}_{\mathrm{L}}^{-1}(H) = L(q^s, \dot{q}^s) = \dot{q}^\alpha p_\alpha (q^s, \dot{q}^s) - H(q^\alpha, p_\alpha (q^s, \dot{q}^s)),
\end{equation}
\emph{in which} \( p_{\alpha} = p_{\alpha}(q^s, \dot{q}^s) \) \text{'s}, \(\alpha = 1, \cdots, n\), \emph{are solutions of the system of equations}
\begin{equation}\label{Eq.(3.39)}
\dot{q}^{\alpha} = \frac{\partial H}{\partial p_{\alpha}}, \quad \alpha = 1, \cdots, n,
\end{equation}
\emph{for the fibres} \(p_{\alpha}\)\text{'s}, \emph{in terms of the dynamical variables} \( q^{\alpha} \)\text{'s} and \( \dot{q}^{\alpha} \)\text{'s}. \emph{This map transforms the Hamiltonian function of the DS into the its Lagrangian function is the inverse of the map} \( \mathbf{\Phi}_{\mathrm{L}} \).

\( \)

The similarity of Eqs. \eqref{Eq.(3.37)}, \eqref{Eq.(3.38)} and \eqref{Eq.(3.39)} in Hamiltonian formulation with Eqs. \eqref{Eq.(3.33)}, \eqref{Eq.(3.34)} and \eqref{Eq.(3.36)}
in Lagrangian formulation is remarkable. Eqs. \eqref{Eq.(3.37)}, \eqref{Eq.(3.38)} and \eqref{Eq.(3.39)} have the same role in the Hamiltonian formulation as
their corresponding Eqs. \eqref{Eq.(3.33)}, \eqref{Eq.(3.34)} and \eqref{Eq.(3.36)} have in Lagrangian formulation. By having a group of them,
for example Eqs. \eqref{Eq.(3.33)}, \eqref{Eq.(3.34)} and \eqref{Eq.(3.37)}, by changing \( L \to H \), \( \dot{q}^{s} \to p_s \), \( \omega_{\alpha\beta} \to \Omega^{\alpha\beta} \), and \( \mathbf{\Phi}_{\mathrm{L}} \to \mathbf{\Phi}_{\mathrm{L}}^{-1} \), one can easily produce the equations of the second groups, Eqs. \eqref{Eq.(3.37)}, \eqref{Eq.(3.38)} and \eqref{Eq.(3.39)}
and vice versa, having the equations of the second category, changing \( H \to L \), \( p_s \to \dot{q}^{s} \), \( \Omega^{\alpha\beta} \to \omega_{\alpha\beta} \), and \( \mathbf{\Phi}_{\mathrm{L}}^{-1} \to \mathbf{\Phi}_{\mathrm{L}} \), one can easily produce the equations of the first category.


\section{\label{app.B} Canonical transformation in Hamilton-Jacobi theory}

In this Appendix, we first introduce the CT in Hamilton-Jacobi theory.
Then, the method of obtaining the general solutions of system of HCEs (Eq. \eqref{Eq4.1}) and HJE (Eq. \eqref{Eq4.14}) is discussed.

In general, suppose the Hamiltonian function of the DS is time dependent such as \( H(t, q, p) \). In Hamiltonian formalism, the focus is on solving the HCEs:
\begin{equation}\label{Eq4.1}
\dot{q} = \frac{\partial H}{\partial p}, \quad \dot{p} = -\frac{\partial H}{\partial q}.
\end{equation}
The HCEs  \eqref{Eq4.1} form a system of two first-order PDEs for the canonical variables \( q \) and \( p \). In Hamilton-Jacobi theory (HJT) with a known Hamiltonian function \( H(t, q, p) \), instead of directly solving the HCE \eqref{Eq4.1}, a transformation called the CT is first defined.

\(\)
\\
\textbf{Definition B.1} (Canonical Transformation). \emph{Suppose that for a DS with one degree of freedom} \( S_1^2 \) \emph{and Hamiltonian function} \( H(t,q,p) \), \( p \) \emph{is the generalized momentum conjugate to the generalized coordinate} \( q \). \emph{Also, suppose that} \( F_2(t,q,P) := S(t,q,P) \) \emph{is an arbitrary function subject to the condition} \( \partial^2 S / \partial q \partial P \neq 0 \).
\emph{The map} \( {\bf \Phi}_{\rm CT} : \mathbb{R}^+ \times T^* \mathbb{Q} \rightarrow \mathbb{R}^+ \times T^* \mathbb{Q} \), \emph{where} \((t,q,p) \mapsto {\bf \Phi}_{\rm CT}(t,q,p) = (t,Q,P)\), \emph{keeps the form of the HCEs} \eqref{Eq4.1} \emph{invariant, that is, in the transformed system} \((t,Q,P)\), \emph{HCEs are in the form} \(\dot{Q} = \partial K / \partial P\) {\it and} \(\dot{P} = -\partial K / \partial Q\),
\emph{which is called a CT with the generating function} \( S(t,q,P) \).
\emph{In the new HCEs,} \( K(t,Q,P) \) \emph{is a new Hamiltonian function of the DS which is defined in terms of the old Hamiltonian function} \( H(t,q,p) \) \emph{as follows} \cite{60,72,73}:
\begin{equation}\label{Eq4.2}
K = H + \frac{\partial S}{\partial t} .
\end{equation}
\emph{In the other words, the invertible transformation} \((t,q,p) \mapsto {\bf \Phi}_{\rm CT}(t,q,p) = (t,Q,P)\) \emph{is said to be a CT if and only if there exist functions} \( K(t,Q,P) \) \emph{and} \( S(t,q,P) \) \emph{such that Eq.} \eqref{Eq4.2} \emph{is satisfied}.

\(\)

For DSs, there are four types of the CTs \cite{64,74}.
The CT mentioned above is the second of these four types.
Since solving the HCEs in the transformed system, \(\dot{Q} = \partial K / \partial P\), \(\dot{P} = -\partial K / \partial Q\), is not easier than solving the old HCE, that is, \(\dot{q} = \partial H / \partial p\), \(\dot{p} = -\partial H / \partial q\), why is the CT used? The reason for using CT is that the generating function \( S(t,q,P) \) is arbitrary. If this function is chosen so that the new Hamiltonian function \( K \) is equal to zero, then by performing a simple integration of the transformed HCEs, \(\dot{Q} = \partial K / \partial P\), \(\dot{P} = -\partial K / \partial Q\), one can obtain the following results: \( Q = c_2 \), and \( P = c_1 \), where \( c_1 \) and \( c_2 \) are the constants of integration. So, the new canonical variables \( P \) and \( Q \) are both fixed valued.

Now, to obtain the CT we reconsider the HCEs \eqref{Eq4.1} in the old canonical coordinates \((t,q,p)\), and the transformed HCEs
\begin{equation}\label{Eq4.3}
\dot{Q}=\partial K/\partial P,\ \ \ \dot{P}=-\partial K/\partial Q,
\end{equation}
in the new canonical coordinates \((t,Q,P)\). The simultaneous validity Eqs. \eqref{Eq4.1} and \eqref{Eq4.3} implies the simultaneous validity of the following variational principles:
\begin{equation}\label{Eq4.4}
\delta\int_{t_{1}}^{t_{2}}\big{(}p\dot{q}-H(t,q,p)\big{)}\mathrm{d}t=0,
\end{equation}
\begin{equation}\label{Eq4.5}
\delta\int_{t_{1}}^{t_{2}}\Big{(}P\dot{Q}-K(t,Q,P)\Big{)}\mathrm{d}t=0.
\end{equation}
Eqs. \eqref{Eq4.4} and \eqref{Eq4.5} will simultaneously hold if the respective integrands differ by the total derivative of an arbitrary function such as \(\Sigma(t,q,p)\). Thus, the transformation \((t,q,p)\mapsto{\bf \Phi}_{\rm CT}(t,q,p)=(t,Q,P)\) keeps the form of the HCEs invariant, if the following equation is satisfied
\begin{equation}\label{Eq4.6}
p\dot{q}-H=P\dot{Q}-K(t,Q,P)+\frac{\mathrm{d}\Sigma}{\mathrm{d}t}.
\end{equation}
Eq. \eqref{Eq4.6} can be rewritten in terms of the differential forms as follows:
\begin{equation}\label{Eq4.7}
p\mathrm{d}q-P\mathrm{d}Q+(K-H)\mathrm{d}t=\mathrm{d}\Sigma\,.
\end{equation}
Eq. \eqref{Eq4.7} is called the ``characteristic equation'' of the CT. The form of this equation suggests considering \(\Sigma\) as a function of the old and the new generalized coordinates. Suppose the second equation of the CT, \(P=P(t,p,q)\) can be solved for \(p\) in terms of \(t\), $q$, and $P$, that is, \(p=p(t,q,P)\). In this case, the second equation of the CT, i.e. \(Q=Q(t,q,p)\) allows us to write the transformed generalized coordinate \(Q\) in terms of the variables \(t\), \(q,P\), that is, \(Q=Q\big{(}t,q,p(t,q,P)\big{)}\). If we take \(q\) and \(P\) as two independent variables, then there exists a generating function of the form
\begin{equation}\label{Eq4.8}
F_{2}(t,q,P):=PQ(t,q,P)+\Sigma\big{(}t,q,p(t,q,P)\big{)}.
\end{equation}
This generating function is of the second type among the four different types for the generating function.
In this function, the variables \(q\) and \(P\)  are selected from the old canonical variables \((t,q,p)\)
and new canonical variables \((t,Q,P)\), respectively.
We obviously have the identity \(-P\mathrm{d}Q=-{\rm d}(PQ)+Q\mathrm{d}P\). By substituting this identity into Eq. \eqref{Eq4.7} one can get
\begin{equation}\label{Eq4.9}
p\mathrm{d}q+Q\mathrm{d}P+(K-H)\mathrm{d}t=\mathrm{d}F_{2}.
\end{equation}
Here we have used Eq. \eqref{Eq4.8}. Notice that \(F_{2}\) is a function of the variables \(t\), \(q\), and \(P\) such that
\begin{equation}\label{Eq4.10}
\mathrm{d}F_{2}(t,q,P)=\frac{\partial F_{2}}{\partial t}\,\mathrm{d}t+\frac{\partial F_{2}}{\partial q}\,\mathrm{d}q+\frac{\partial F_{2}}{\partial P}\,\mathrm{d}P.
\end{equation}
Plugging Eq. \eqref{Eq4.10} into Eq. \eqref{Eq4.9} yields
\begin{equation}\label{Eq4.11}
\Big(K-H-\frac{\partial F_{2}}{\partial t}\Big)\mathrm{d}t+\Big(p-\frac{\partial F_{2}}{\partial q}\Big)\mathrm{d}q+\Big(Q-\frac{\partial F_{2}}{\partial P}\Big)\mathrm{d}P=0.
\end{equation}
Since \(t\), \(q\) and \(P\) are independent variables, Eq. \eqref{Eq4.11} always holds if and only if the coefficients of \(\mathrm{d}t\), \(\mathrm{d}q\) and \(\mathrm{d}P\) are independently equal to zero,
i.e.,
\begin{equation}\label{Eq4.12}
p=\frac{\partial F_{2}}{\partial q},\quad~~~~ Q=\frac{\partial F_{2}}{\partial P}, ~~~~~~~K-H=\frac{\partial F_{2}}{\partial t}.
\end{equation}
Eqs. \eqref{Eq4.12} give two PDEs for the two unknowns, \(P(t,q,p)\) and \(Q(t,p,q)\).
To find an explicit form for the CT, by taking the partial derivative of \(F_{2}\) with respect to variable \(q\) we obtain
an algebraic equation in terms of the variables \(t\), \(q\), \(p\) and \(P\). This equation is  \(Z:=p-Y(t,q,P)=0\) where \(Y(t,q,P):=\partial F_{2}(t,q,P)/\partial q\).
Then, by solving this algebraic equation for the variable \(P\), one can obtain \(P=P(t,q,p)\) which is the first equation of the CT. Now, by inserting this result into second equation of \eqref{Eq4.12}
yields the following algebraic equation
\[\frac{\partial F_{2}(t,q,P)}{\partial P}:=\omega(t,q,P)=\omega\big{(}t,q,P(t,q,p)\big{)}:=Q(t,q,p),\]
which is the second equation of the CT \eqref{Eq4.12}. In this way, the system of equations \(P=P(t,q,p)\) and \(Q=Q(t,q,p)\) specify the CT. Under this transformation, the transformed Hamiltonian function \(K(t,q,P)\) and the old Hamiltonian function \(H(t,q,p)\) are related by the third equation of \eqref{Eq4.12}.
As mentioned earlier in Definition B.1, we denoted this generating function by \(S(q,P,t)\), that is \(F_{2}(t,q,P):=S(t,q,P)\). Now,
by substituting \(K=0\) and \(p=\partial S/\partial q\) into the third equation of \eqref{Eq4.12} we obtain
\begin{equation}\label{Eq4.14}
H\Big(t,q,\frac{\partial S}{\partial q}\Big)+\frac{\partial S}{\partial t}=0.
\end{equation}
This is a PDE of the first-order in time \(t\) and generalized coordinate \(q\) for the generating function \(S(t,q,P)\), which is called the HJE.
In this equation, the generating function \(S\) is usually called the Hamilton's principal function \cite{74}. Therefore, the solution of the HCEs \eqref{Eq4.1} is equivalent to the solution of the first-order PDE \eqref{Eq4.14}. Solving Eq. \eqref{Eq4.14} is much easier than solving a system of the PDEs. Any solution \(S\) of the HJE \eqref{Eq4.14} satisfying the Hessian condition \(\partial^{2}S/\partial q\partial P\neq 0\), is called a complete solution. Historically, for the first time, W. R. Hamilton derived Eq. \eqref{Eq4.14} in 1834.
Later in 1837, J. Jacobi made a precise connection between the solutions of Eq. \eqref{Eq4.14} and his studies lead to a theorem which is called the Hamilton-Jacobi Theorem.
\(\)
\\
\textbf{Theorem B.1} (Hamilton-Jacobi, 1837). \emph{Suppose that the function} \(S(t,q,P)\) \emph{is a complete solution of the HJE} \eqref{Eq4.14}. \emph{Then the general solution of the HCEs} \eqref{Eq4.1} \emph{is given by the system of PDEs}
\begin{equation}\label{Eq4.15}
\frac{\partial S(t,q,P)}{\partial P}=Q,\quad\frac{\partial S(t,q,P)}{\partial q}=p,
\end{equation}
\emph{where} \(Q\) \emph{is an arbitrary constant} \cite{3}.
\\
\textbf{Proof.} See Ref. \cite{3}.

\(\)

Today, it is known that the HJT has applications in the gravity and cosmology. This theory has a key role in solving the gravitational field equations. The HJT provides a starting point for a semiclassical analysis used in stochastic inflation \cite{75,76}, and also in quantum cosmology. For the first time, in 1962, HJE was used by A. Peres \cite{77}. In their research, Peres {\it et al.} were trying to formulate the theory of GR to include quantum theory. Considering the Einstein-Hilbert action and applying the principle of least action and the Arnowitt-Deser-Misner formalism, Peres found the tensor form of the HJE in curved spacetime.
This equation is a generalized form of the HJE in curved Riemaniann spacetime and is known as the Hamilton-Jacobi-Einstein equation (HJEE).
In 1994, J. Parry {\it et al.} \cite{75} by including matter fields presented a systematic method for solving the HJE. This method can be used to derive the ``Zeldovich approximation'' in the theory of GR. In recent years, systematic and practical methods for calculating the generating function in cosmology have been proposed by many authors \cite{76,78,79,80,81,82,83}. What we do in this study is to find a CT that keeps the forms of the HCEs invariant. By using this transformation, then we obtain the generating function of the CT. Our goal in doing this is to find two independent invariants (first integrals) for the FLRW cosmological model whose Hamiltonian function is known. For this purpose, we return to our problem. Once the complete solution of the HJE is obtained,
the first two equations of \eqref{Eq4.12} give the following functions \cite{61,72,74,81}
\begin{equation}\label{Eq4.16}
\alpha(t,q,p,P):=p-\frac{\partial S(t,q,P)}{\partial q}=0,
\end{equation}
\begin{equation}\label{Eq4.17}
\beta(t,q,Q,P):=Q-\frac{\partial S(t,q,P)}{\partial P}=0,
\end{equation}
where \(\alpha\) and \(\beta\) are known functions of the variables (\(t\), \(q\), \(p\), \(P\)) and  (\(t\), \(q\), \(Q\),\(P\)), respectively.
Now, we consider Eqs. \eqref{Eq4.16} and \eqref{Eq4.17} as a system of algebraic equations and solve them simultaneously for the canonical variables \(Q\) and \(P\) in terms of the other variables \(t\), \(p\) and \(q\):
\begin{equation}\label{Eq 4.18}
Q=Q(t,q,p),\quad P=P(t,q,p).
\end{equation}
In fact, these equations are the CT produced by the generating function \(S(q,P,t)\). Since \(P\) and \(Q\) are both constants of motion, then, we will have the following two independent invariants:
\begin{equation}\label{Eq4.19}
P(t,q,p)=c_{1},\ \ Q(t,q,p)=c_{2}\,
\end{equation}
where \(c_{1}\) and \(c_{2}\) are constants. In the final step, using Eq. \eqref{Eq.(3.36)} we find
\begin{equation}\label{Eq4.20}
p=\frac{\partial L(t,q,\dot{q})}{\partial\dot{q}}:=\chi(t,q,\dot{q}).
\end{equation}
By substituting the function \(\chi(t,q,\dot{q})\) instead of the variable \(p\) in
Eq. \eqref{Eq4.19} one then obtains
\begin{equation}\label{Eq4.21}
I_{\text{HJ},1}(t,q,\dot{q}):=P(t,q,\chi(t,q,\dot{q}))=c_{1},
\end{equation}
\begin{equation}\label{Eq4.22}
I_{\text{HJ},2}(t,q,\dot{q}):=Q(t,q,\chi(t,q,\dot{q}))=c_{2}.
\end{equation}
Therefore, solving the HJE  \eqref{Eq4.14} leads us to the extraction of two independent invariants (or first integrals), which we call in this study the IHJ. By simultaneously solving these invariants as a system of algebraic equations, the general solution of the DS can be obtained. To solve the system of equations \eqref{Eq4.21} and \eqref{Eq4.22}, first we solve one of these invariants, for example, the first invariant  \eqref{Eq4.21} for the variable \(\dot{q}\) in terms of the other variables \(t,q\) and the constant of motion \(c_{1}\) as follows: \(\dot{q}=\Omega(t,q;c_{1})\), where \(\Omega\) is a known function of the variables \(t\), \(q\) and the constant of motion \(c_{1}\). By substituting it in the second invariant, \eqref{Eq4.22}, then we will have: \(I_{\text{HJ},2}\big{(}t,q,\Omega(t,q;c_{1})\big{)}=c_{2}\). Finally, by solving this algebraic equation for the generalized coordinate \(q\) in terms of the time \(t\) and constants of motion \(c_{1}\) and \(c_{2}\), the final solution is obtained as follows: \(q=q(t;\ c_{1},c_{2})\), which is the general solution of HJE  \eqref{Eq4.14}.
Indeed, this solution is also the general solution of the system of the HCEs \eqref{Eq4.1}.


\end{document}